\documentclass{aa} 
\usepackage{booktabs}
\usepackage{graphicx}

\newcommand{\orcid}[1]{\href{https://orcid.org/#1}{\includegraphics[width=8pt]{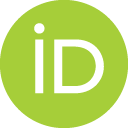}}}
\usepackage{txfonts}
\usepackage{lipsum}
\usepackage{subcaption}         % necessary for continued figures, example in section 3
\usepackage[colorlinks=true,linkcolor=blue,citecolor=blue]{hyperref}
\usepackage{lscape}             % to rotate a single page table, example in appendix.
\usepackage{placeins}           % useful with \FloatBarrier, to keep 
\begin{document}

%%%%%%%%%%%%%%%%%%%%%%%%%%%%%%%%%%%%%%%%
% if you use custom commands in your title,
% ensure to check your title when submitting!
%%%%%%%%%%%%%%%%%%%%%%%%%%%%%%%%%%%%%%%%

\title{Exploring the capabilities of combined modelling strategies for supernova with relativistic jets using \textsc{Redback}}
   % \title{An Investigation of Fitting Strategies for GRB-SN events using Redback}

% Exploring the capabilities of combined modelling strategies for GRB-SN events with Redback

% Investigation of a new approach to the modelling of GRB-SNe

% Moving towards population level studies of GRB-SNe: an investigation into joint modelling cababilities of GRB-SN events with Redback

%%%%%%%%%%%%%%%%%%%%%%%%%%%%%%%%%%%%%%%%
% Please separate each author with the \and command
%
% Please do not include ORCIDs next to author names.
% Only ORCIDs authenticated by individual authors in EDPS
% editorial system will be taken into account.
% ORCIDs included here will be removed.
%%%%%%%%%%%%%%%%%%%%%%%%%%%%%%%%%%%%%%%%

   \author{L. Cotter \inst{1}\fnmsep\thanks{Corresponding author: laura.c.cotter@gmail.com}\orcid{0000-0002-7910-6646}
        \and N. Sarin\inst{2,3}\orcid{0000-0003-2700-1030}
        \and A. Martin-Carrillo\inst{1}\orcid{0000-0001-5108-0627}
        }

\institute{School of Physics and Centre for Space Research, University College Dublin, Belfield D04 V1W8, Dublin, Ireland 
    \and
    Kavli Institute for Cosmology, University of Cambridge, Madingley Road, CB3 0HA, UK
    \and 
    Institute of Astronomy, University of Cambridge, Madingley Road, CB3 0HA, UK
    }

   \date{Received xxxx; accepted xxxx}

\titlerunning{Fitting Strategies for SNe with relativistic jets}
\authorrunning{Cotter et al. }

\abstract{
 Events in which a supernova (SN) is associated with a relativistic jet present multi-component light curves. In the past, the analysis of such complex light curves was carried out largely by investigating each component individually due to limitations in modelling and inference. Such decoupled analyses are susceptible to biased parameter estimation due to cross-contamination between the emission of the relativistic jet and the SN. Additionally, differences in adopted methodologies make it difficult to compare physical parameters across populations. In this work, we investigate three common modelling strategies adopted for events where a  Gamma-ray Burst (GRB) is associated with an SN and compare them with a joint model and inference approach. We assess the ability of each approach to fit and reproduce the true injection parameters of 1000 simulated GRB-SN datasets using \textsc{Redback}. We split these datasets into three common GRB-SN light curve morphologies: one where the SN is dominant, one where the GRB is dominant, and one where the SN and GRB have relatively equal contributions to the light curve. We also investigate the behaviour of each modelling technique when the jet break of the GRB afterglow occurs at early times and at late times. We find that treating each component independently with an individual model leads to significant bias in the estimated parameters even though the resulting best fit to the data may appear acceptable. We further demonstrate that afterglow subtraction is unreliable for GRB-SN modelling as uncertainties in the afterglow fit are not fully propagated into the resulting afterglow-subtracted light curve.   Instead, our results show that jointly modelling the relativistic jet and SN components provides the most reliable modelling approach, particularly when multi-band X-ray and radio observations are available.  }


\maketitle

%%%%%%%%%%%%%%%%%%%%%%%%%%%%%%%%%%%%%%%%%%%%%%%%%%%%%%%%%%%%%%
\section{Introduction}

The discovery of the first Gamma-ray Burst (GRB) associated with a supernova (SN), GRB\,980325/SN\,1998bw, marked the first concrete link between SNe and relativistic jets \citep{Gal1998bw}. Since then, the GRB-SN population has grown substantially, now comprising over 60 well-observed events \citep{finneran_2025_webtool}. In recent years, with the launch of the \textit{Einstein Probe (EP)} mission, there has been an influx of detections of Fast X-ray transients (FXTs) associated with SNe Ic-BL. The link between long GRBs and SNe Ic-BL is well established, with both transients arising from the core collapse of massive stars, releasing comparable isotropic energies of order $10^{52}$ erg and $10^{51}$ erg, respectively \citep{woosley2006supernova,hjorth2012grb,cano2017observer}. In contrast, the origins of FXTs remain unclear, with these events thought to encompass a broader range of transient phenomena such as tidal disruption events (TDEs) \citep{Jonker2013,glennie2015},
bursts originating from the rapid spin-down of newly born millisecond magnetars produced by neutron star
mergers \citep{Zhang_magnetar,2024Vasquez} and unusual GRBs \citep{Sarin2021, 2024_Wichern}. FXTs associated with SNe Ic-BL are thought to arise from several physical mechanisms, including SN shock breakout \citep[e.g. SN\,2008D;][]{2008D}, jets choked within the progenitor star \citep[e.g. EP260321a/SN\,2026gzf;][]{martincarrillo2026,oconnor2026,rastinejad2026,chen2026,yuan2026}, or intrinsically low-luminosity GRBs \citep[e.g. EP250304a/SN\,2025fhm, EP250827b/SN\,2025fhm][Corcoran et al. 2026, \textit{in prep}]{cotter2026n,0827b}. Of course, relativistic jets associated with SNe can also be viewed off-axis, where the observer's line of sight lies outside the initial opening angle of the jet. In the case of GRBs, the highly relativistic emission is strongly beamed away from the observer such that the GRB prompt emission cannot be observed. As the jet decelerates, it spreads into the observer's line of sight, where the resulting afterglow appears at much later times than for an on-axis GRB, potentially appearing as an FXT rather than a classical GRB \citep{Sarin2021, 2024_Wichern,2026_zheng}.

As the number of these diverse FXT/GRB-SNe with rich multi-band observations continues to grow, astronomers will soon be able to perform population studies and more detailed modelling of such events, providing new constraints on progenitor properties and explosion mechanisms and allowing for a deeper investigation into the diversity of these relativistic jets. However, performing modelling of dual-component transient light curves is far from straightforward. Many difficulties arise when attempting to perform multi-band modelling of FXT/GRB-SN light curves due to cross-contamination between the emission components. In the case of GRB-SN events, early-time emission is dominated by synchrotron emission from the GRB afterglow, which is well described by a power-law decay \citep{1998_sari_piran}. At later times, the SN component emerges as a broad, thermal peak in the light curve \citep{arnett1982}. The temporal overlap of these components complicates data interpretation and introduces degeneracies in model fitting. The growing sample of FXT-SNe are potentially even more challenging to model, as the physical origin of the early-time emission is not always well established. Even ignoring the inference perspective, self-consistently modelling the interplay between the jet and ejecta is not well captured by many models.

As FXT-SNe are a relatively new population, detailed modelling of their multi-band emission remains limited. In contrast, several approaches have been used in the past to disentangle the overlapping components of GRB-SNe light curves. One such method involves fitting the early data with an afterglow model and the late-time data with the SN model.  This method has limitations as it does not account for the late-time afterglow behaviour, when the SN emission becomes dominant. This could result in the fitted SN data still having some contamination from the afterglow, leading to an apparently brighter peak and an overestimation of the mass of nickel present in the ejecta. It is also difficult to determine exactly what is pure afterglow and what is SN emission, which can lead to the afterglow-only fit containing some SN contamination. The second method involves fitting a broken power-law or afterglow model to the beginning of the light curve where the GRB component is dominant and then subtracting the best-fitting model such that only the SN emission remains \citep[e.g.][]{2004_Zeh, Fulton_2023,belkin_2024}. Again, similar issues in parameter estimation arise, as errors from the afterglow model that are subtracted from the data are typically not properly accounted for when modelling the afterglow-subtracted SN. Moreover, this ultimately conditions inferred SN parameters on a specific model and best-fitting set of parameters of the afterglow, somewhat defeating the value of inference: marginalisation. 

In this paper, we focus on GRB-SN events, where we use a combined model in \texttt{Redback}~\citep{Redback} using the \texttt{tophat\_from\_emulator} model for the afterglow component, based on the modelling outlined in \citet{lamb_2018} and \citet{Wallace_2025}, and the classical SN model proposed by \citet{arnett1982} and perform inference on simulated light curves to understand the systematics associated with different inference strategies for GRB-SN. In particular, we explore the feasibility of this model in accurately fitting GRB-SN events by creating a variety of simulated GRB-SN datasets. We test our joint fit methodology and compare results with other modelling methodologies to determine which method best reproduces the injected parameters.

\section{Methodology}

\subsection{Modelling}
GRB-SN light curves are inherently complex, with multiple physical components contributing to the observed emission at different epochs and wavelengths. In this work, we adopt a simplified framework consisting of a tophat GRB afterglow model (\texttt{tophat\_from\_emulator}) and an Arnett SN model, whose flux contributions are summed to produce a combined GRB-SN light curve. For all fitting, we make use of the \texttt{nessai} nested sampler \citep{nessai}, which is implemented in \textsc{Redback} through the \textsc {BILBY} framework \citep{bilby}.

The afterglow component is described by the \texttt{tophat\_from\_emulator} model, which is based on the work of \citet{lamb_2018} and \citet{Wallace_2025}. The model assumes an ultra-relativistic, narrow, tophat jet propagating into a uniform-density interstellar medium (ISM). To reduce computational cost, the model is evaluated using a multilayer perceptron (MLP) regressor trained on a grid of afterglow simulations. Additional simplifications include fixing the redshift and assuming that all electrons swept up by the shock are accelerated (ksin $\approx 1$). These assumptions are appropriate for the present study, which focuses on assessing parameter recovery across a large ensemble of simulated transients rather than performing detailed modelling of individual events. The free parameters of the \texttt{tophat\_from\_emulator} afterglow model are the isotropic-equivalent kinetic energy ($E_{\rm k, iso}$), the observer viewing angle ($\theta_v$), the jet half-opening angle ($\theta_c$), the ambient interstellar medium density ($n_{\rm ism}$), the electron power-law index (p), the electron energy fraction ($\epsilon_e$), the magnetic energy fraction ($\epsilon_B$), and the initial Lorentz factor ($\Gamma_0$).

The supernova component is described by the one-zone Arnett model, in which the optical luminosity is powered by the radioactive decay of $^{56}$Ni to $^{56}$Co \citep{arnett1982}. The shape and evolution of the light curve are determined primarily by the ejecta mass ($m_\text{ej}$), ejecta velocity ($v_\text{ej}$ and the mass of $^{56}$Ni, which together govern the diffusion timescale and radioactive energy deposition. The model further assumes homologous expansion, spherical symmetry, centrally concentrated $^{56}$Ni, and a constant opacity ($\kappa$). While these assumptions are known to be oversimplified for SNe Ic-BL and other stripped envelope SNe, the Arnett model remains widely used owing to its computational efficiency and ability to recover first-order explosion properties. In GRB-SN explosions, asymmetric jets can drive hydrodynamic instabilities as they propagate through the progenitor, mixing  $^{56}$Ni to larger radii and, in some cases, transporting metal-rich material ahead of the bulk ejecta \citep{Woosley_2002,barnes_2018,sarin2026}. This outward mixing can lead to a faster rise in the light curve than predicted by spherically symmetric models \citep{dessart_2012,ashall2019}. Consequently, the Arnett model should be viewed as an approximate description of the supernova emission rather than a complete physical representation of the explosion. Nevertheless, its simplicity makes it well suited for investigating the performance of simultaneous GRB-SN fitting within a controlled simulation framework.

\subsection{Simulations}\label{sim}
To truly test the model's capabilities, we first create a simulated dataset with known, realistic physical parameters. To ensure that the parameters selected for the simulation are representative of a GRB-SN, we first investigate the properties of the SN Ic-BL and GRB-SN populations. \citet{Taddia_2019} showed that the average absolute peak magnitude of normal SNe Ic-BL is about -18.6 in the \textit{r} band. The absolute peak magnitude of typical GRB-SNe in the same band may be slightly higher, with the maximum recorded belonging to SN\,2011kl at a peak absolute magnitude of -20 in the V band \citep{cano2017observer}. \citet{Taddia_2019} also showed in their population studies that the kinetic energies of SNe Ic-BL lie between $10^{51}$ and $10^{52}$ ergs. The mass of $^{56}$Ni synthesised by SNe associated with GRBs has been observed to be higher than the typical SN Ic-BL population. \citet{cano2017observer} provides a list of $^{56}$Ni masses for a sample of the GRB-SN population, with the reasonable maximum being 0.7 $M_\odot$ and a maximum nickel fraction of 0.25. Given this information, we ensure that the random simulation of parameters must meet 4 selection criteria (1) the peak absolute magnitude of the simulated SNe must reflect the population of GRB-SNe/ SNe Ic-BL, (2) the kinetic energy of the SN much be within a reasonable range of $ \sim 10^{51} - 10^{52}$ ergs, (3) the nickel mass must be less than 0.7 $M_\odot$ and (4) the fraction of nickel in the ejecta mass must be less than 0.25.\footnote{The kinetic energy of the SN is calculated using the typical formula $E_\text{ke, sn} = \frac{1}{3} \,m_\text{ej}\,v_\text{ej}^2$}. These conditions enabled us to choose realistic initial parameters, shown in Table \ref{parameters}.

When simulating the GRB component of the light curves, we had similar criteria to ensure that the simulated transients were realistic. The simplified \texttt{tophat\_from\_emulator} model struggles to create early-time light curves when the jet break is too early. To combat this, we set a constraint on the jet break time, which is calculated using the formula provided by \citet{2009_racusin} with the simulated GRB parameters. Any parameter set that returned a break time of less than 0.3 days was rejected. Another constraint that we set is that the microphysical parameter $\text{log}{10}\epsilon_e$ is always greater than $\text{log}{10}\epsilon_b$ \citep{2007_Medved}. We allow the modelling to explore a set range for the $\theta_\text{c}$, $\Gamma_0$ and temperature floor parameters; however, to simplify the simulation process and ensure that the simulated data sets remained realistic, we set these parameters to the same fixed values for each simulation, as shown in Table~\ref{parameters}.

\begin{table}[h]
\centering
\small
\setlength{\tabcolsep}{3pt}
\caption{Parameters used to simulate the GRB-SN datasets.}
\renewcommand{\arraystretch}{1.1}

\begin{tabular}{l c | l c}

    \toprule
    Parameter & Range & Parameter & Range \\
    \midrule
    log$_{10}$\,$E_{k \text{iso}}$ (erg) & (49 – 53) & z & 0.1 \\
    $\theta_\text{v}$ (radians) & 0 & $A_v$ & 0 \\
    $\theta_\text{c}$ (radians) & 0.1 & $m_\text{ej}$ ($M_\odot$) & (1.5 – 8) \\
    log$_{10}$\,$n\text{ism}$ (g,cm$^{-3}$) & (-2 – 0) & $v_\text{ej}$ (km,s$^{-1}$) & (15000 – 25000) \\
    \textit{p} & (2.2 – 2.5) & f\_nickel & (0.01 – 0.2) \\
    log$_{10}$\,$\epsilon_e$ & (-2 – -0.5) & $\kappa$ (cm$^2$ g$^{-1}$) & 0.07 \\
    log$_{10}$\,$\epsilon_B$ & (-4 – -2) & $\kappa_\gamma$ (cm$^2$ g$^{-1}$) & 0.03 \\
    $\Gamma_0$ & 400 & Temp. Floor (K) & 4000 \\

\bottomrule
\end{tabular}

\label{parameters}

\end{table}

The \texttt{Redback} software has several built-in functionalities which allow for the simulation of electromagnetic transients given certain conditions and desired models. To simulate a generic transient, the \texttt{SimulateGenericTransient} function was used. Using our desired model, we simulated 500 data points total in the optical \textit{griz} bands, 500 data points total in the 1.4GHz, 5 GHz, and 15GHz radio bands and 500 data points in the 1 keV X-ray band. We added 30\% Gaussian noise to the data to make it more realistic. Using this full dataset, we then created the transient light curve by randomly drawing data points from this larger dataset. For the optical bands, the simulated dataset included 30 points in the \textit{r}-band, 20 in the \textit{g}-band and 15 in both the \textit{i}-band and \textit{z}-bands, totalling 80 data points across the optical bands. 5 data points were drawn for each radio band, and 20 data points were drawn in the 1 keV X-ray band. Data points drawn within 10\% of one another in time within a given filter were subsequently combined.

\begin{figure}[h!]
\centering
\includegraphics[width=0.9\linewidth]{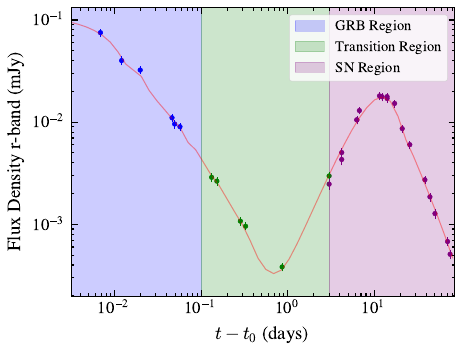}
\caption{Example of simulated data points for the GRB-SN light curve in each defined region. }
\label{sim_data_regions}
\end{figure}

Since the selection of data was completely random, there were cases where data did not cover certain points in the light curve. We define three key regions where there must be data points: (1) the GRB region, (2) the transition region and (3) the SN region. Drawing from previously studied GRB-SN events that had good data coverage, we estimate that approximately 20\%of the total data points are in the region of the light curve that is GRB-dominated, 20\%are in the transition from GRB-dominated to SN-dominated, and 60\%are in the SN-dominated part of the light curve. We define the GRB region as between 0.0001 days and 0.1 days, the transition region from 0.1 to 3 days and the SN region dominates from 3 days to 80 days. The transition region was difficult to determine, as the turning point created in this region of the light curve can shift forwards or backwards in time depending on whether the SN or the GRB emission is more dominant. To determine the best time range, we looked at the distribution of the times of these minima in the light curve produced from each set of parameters. We found that 90\%of these times lie between 0.1 and 3 days, which we take as our transition region. An example light curve showing these regions can be found in Figure~\ref{sim_data_regions}.

\subsection{Modelling Implementation}

Modelling was implemented within the \textsc{Redback} framework by creating our custom joint \texttt{tophat\_from\_emulator\_and\_arnett} model, which is composed of the \texttt{tophat\_from\_emulator} model and the Arnett model added together. For each parameter, we chose priors over slightly larger ranges than the simulated data to see where the model may explore. These prior ranges are shown in Table \ref{priors}.

\begin{table}[h]
\centering
\small
\setlength{\tabcolsep}{3pt}
\renewcommand{\arraystretch}{1.1}
\caption{Parameter prior ranges used in the fitting of the simulated GRB-SN dataset.}
\begin{tabular}{l c | l c}

\toprule
Parameter & Range & Parameter & Range\\
\midrule
log$_{10}$\,$E_{k, \text{iso}}$  (erg) & (48 – 56) & z & fixed \\
$\theta_\text{v}$ (radians)   &     (0 – 0.3) & $A_v$ & fixed \\
$\theta_\text{core}$ (radians)   &    (0.01 – 0.3) &$m_\text{ej}$ ($M_\odot$) & (0.1 – 10)\\
log$_{10}$\,$n\text{ism}$ (g,cm$^{-3}$) & (-5 – 2)   &$v_\text{ej}$ (km,s$^{-1}$) & (15000 – 35000)\\
\textit{p}  & (2 – 3)  &f\_nickel& (0.05 – 0.4)\\
log$_{10}$\,$\epsilon_e$  & (-5 – 0) & $\kappa$ ($\text{cm}^2 \text{g}^{-1}$) & fixed\\
log$_{10}$\, $\epsilon_B$      & (-5 – 0) & $\kappa_\gamma$ ($\text{cm}^2 \text{g}^{-1}$)& fixed\\
$\Gamma_0$       &   (10 – 2000) &     Temp. Floor (K) & (800 – 8000)\\
\bottomrule
\end{tabular}

\label{priors}

\end{table}

\begin{figure}[h]
\centering

\begin{subfigure}{\linewidth}
\centering
\includegraphics[width=0.9\linewidth]{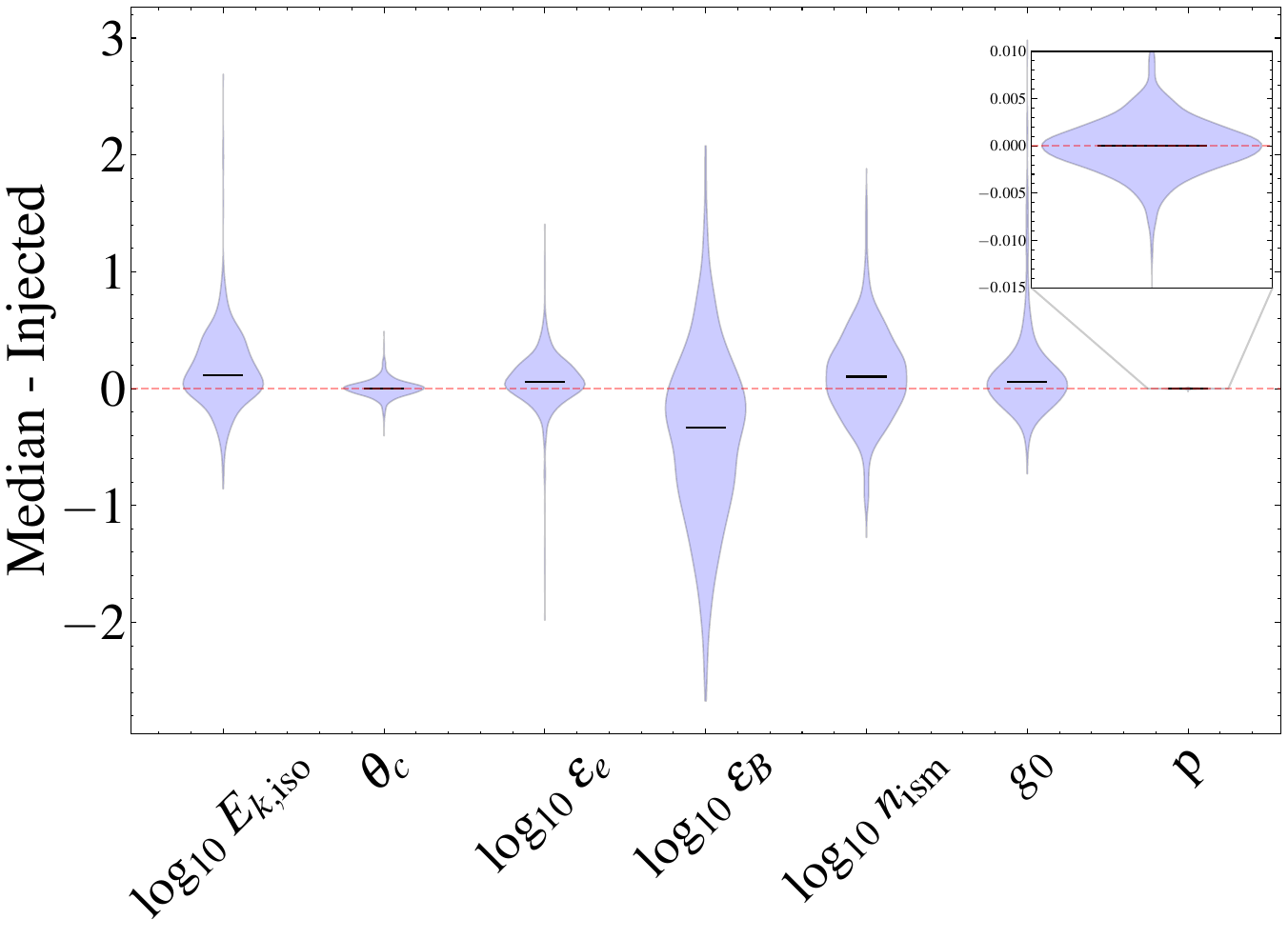}
\caption{\footnotesize Afterglow-only simulation fits.}
\label{grb_vp}
\end{subfigure}

\hfill

\begin{subfigure}{\linewidth}
\centering
\includegraphics[width=0.9\linewidth]{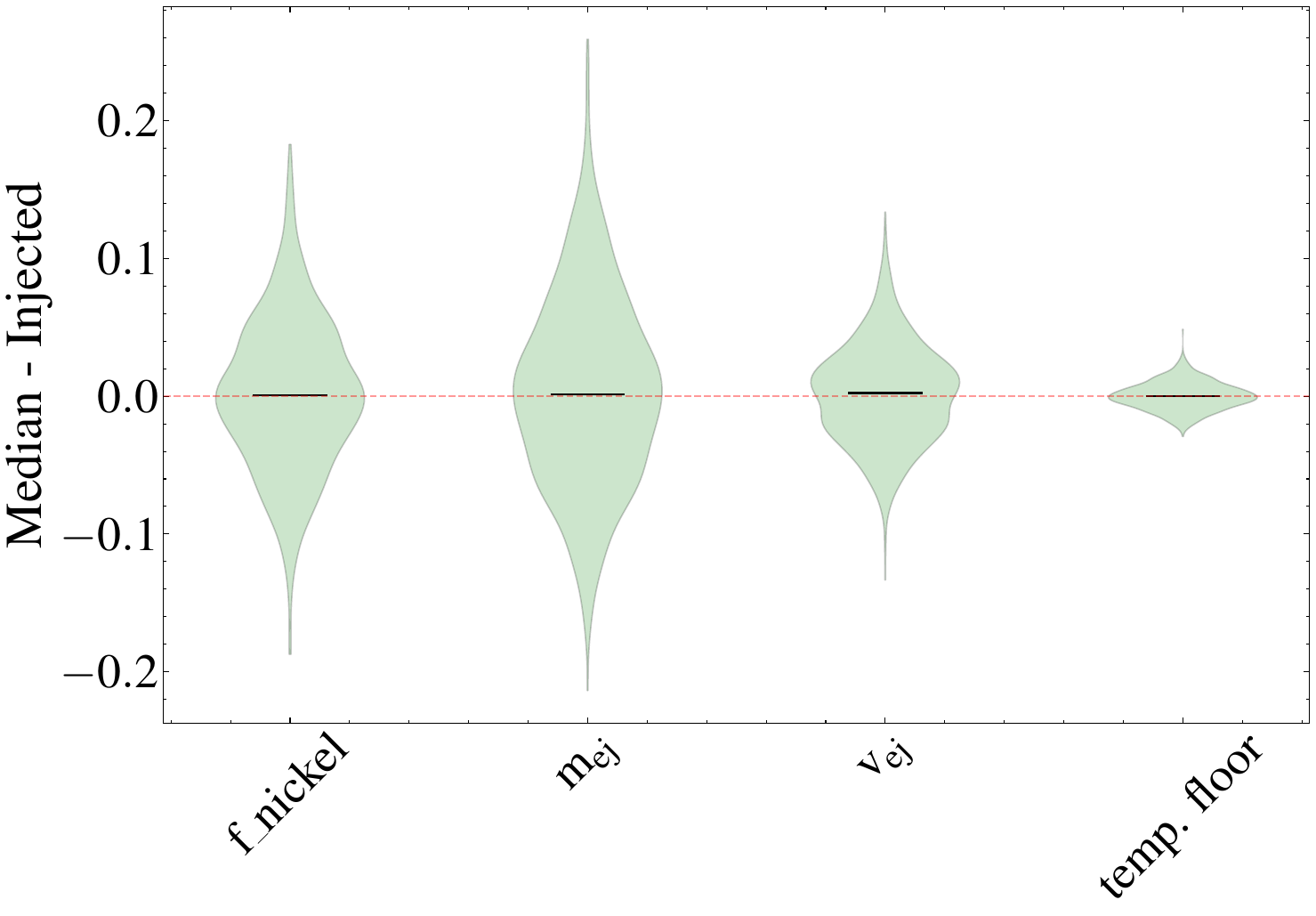}
\caption{\footnotesize SN-only simulation fits.}
\label{sn_vp}
\end{subfigure}

\caption{Violin plots showing the difference between the median posterior of the fits and the injected simulation parameters. The dashed line marks zero difference. }
\label{violin_sep}

\end{figure}

Before performing joint modelling, we created separate GRB and SN datasets to inspect how well the individual \texttt{tophat\_from\_emmulator} and Arnett models work. In each model, there are known degeneracies between different parameters and limitations which could affect the modelling of data \citep{dessart_2016,2024_grb_deg}. Thus, we perform fitting of 1000 simulated SN transients and 1000 simulated GRB transients with these models to investigate whether these degeneracies cause any anomalous behaviour. The simulated datasets were produced in the same way as in Section \ref{sim}, with the SN data comprising only the optical bands and the GRB data consisting of optical, X-ray and radio bands. The priors used in the fitting were kept the same as those shown in Table \ref{parameters}.

In this work, we gauge the performance of the model on the basis of how close the median of the posterior returned by the modelling is to the true value of the simulated parameter or the injected parameter. For each simulated data set fit, we calculate the difference between the median of the posterior and the injected parameter, and then normalise this value by the standard deviation of the posterior sample. The closer this value is to 0, the better the model is at reproducing the light curve. Figure~\ref{violin_sep} shows the distribution of these values for the GRB and SN model tests. In all cases, these distributions are centred close to zero, indicating that the inference procedure is largely unbiased and successfully recovers the injected parameters the majority of the time. Small systematic offsets are observed for the log$_{10}$\,$E_{k, \text{iso}}$, log$_{10}$\,$\epsilon_e$ and log$_{10}$\,$\epsilon_B$ parameters in the GRB fitting run. These offsets are consistent with known parameter degeneracies within the tophat GRB model, where multiple parameter combinations can produce similar light curve behaviour \citep{Garcia_2024}. This can lead to leading to broadened or shifted posterior distributions without implying a failure of the model.

\section{Results}

\subsection{Recovery of GRB-SN parameters across all 1000 simulated datasets}

\begin{figure}[h]
\centering
\includegraphics[width=\linewidth]{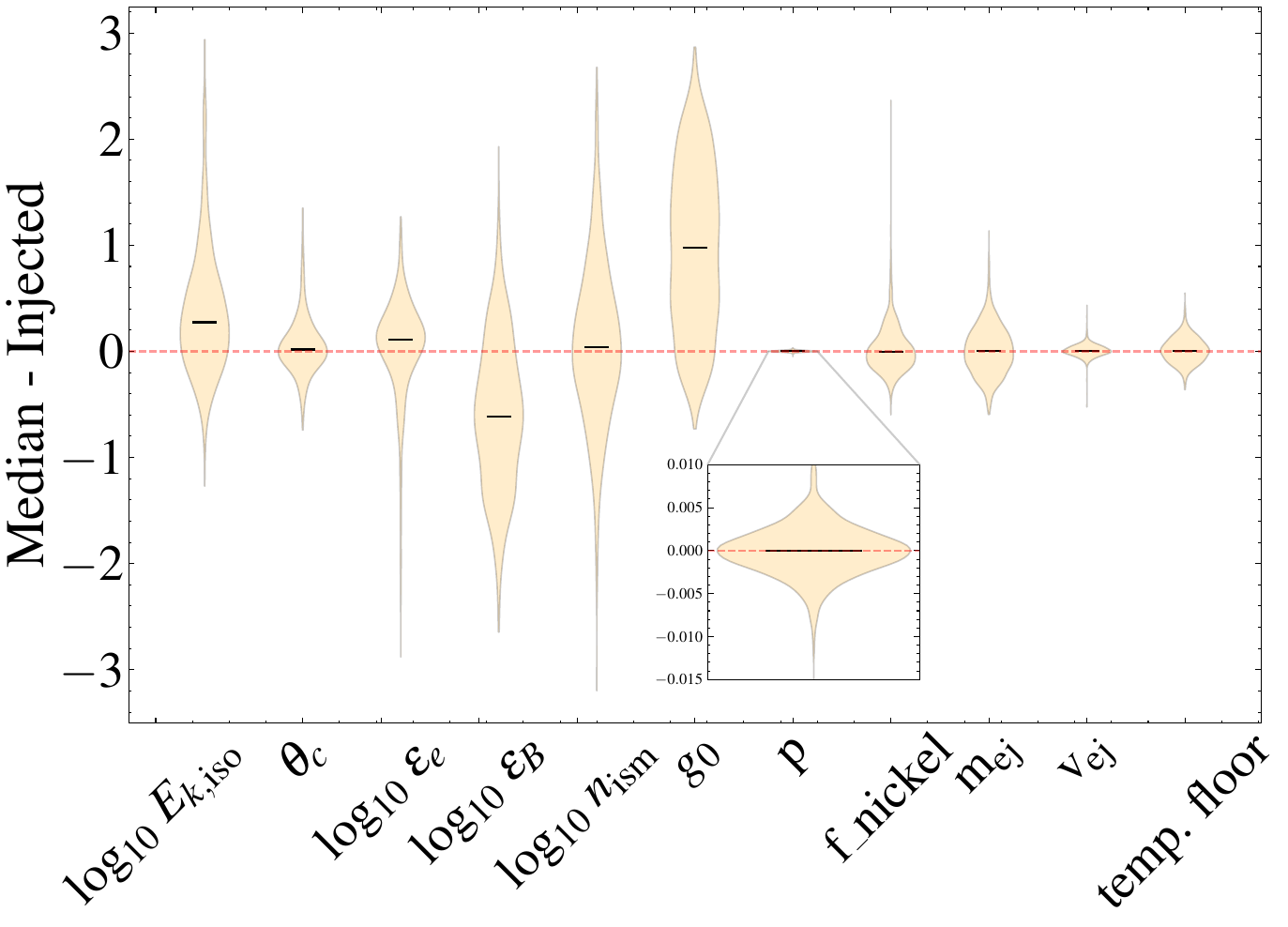}

\caption{Violin plots showing the difference between the median posterior of the fits and the injected simulation parameters for the GRB-SN combined fitting. The dashed line marks zero difference. }
\label{grbsn_vp}

\end{figure}

After investigating the ability of the individual models to reproduce GRB and SN data, we then performed the joint fitting of the GRB-SN data. The results of the fitting of 1000 simulated data sets can be found in Figure~\ref{grbsn_vp}.

The majority of the parameters are centred along 0; however, similar to the results of the GRB tophat model seen in Figure~\ref{grb_vp}, there are offsets from 0 in the case of the log$_{10}$\,$E_{k, \text{iso}}$, log$_{10}$\,$\epsilon_e$, log$_{10}$\,$\epsilon_B$ and $g_0$ parameters. Again, we emphasise that this deviation is not a failure of the model but arises as a result of degeneracies between parameters within the model. Thus, we conclude from the 1000 simulation fits that the joint model can reliably recover a GRB-SN light curve.

\subsection{Simulated dataset behaviour}

\begin{figure}[t!]
\centering
\includegraphics[width=0.9\linewidth]{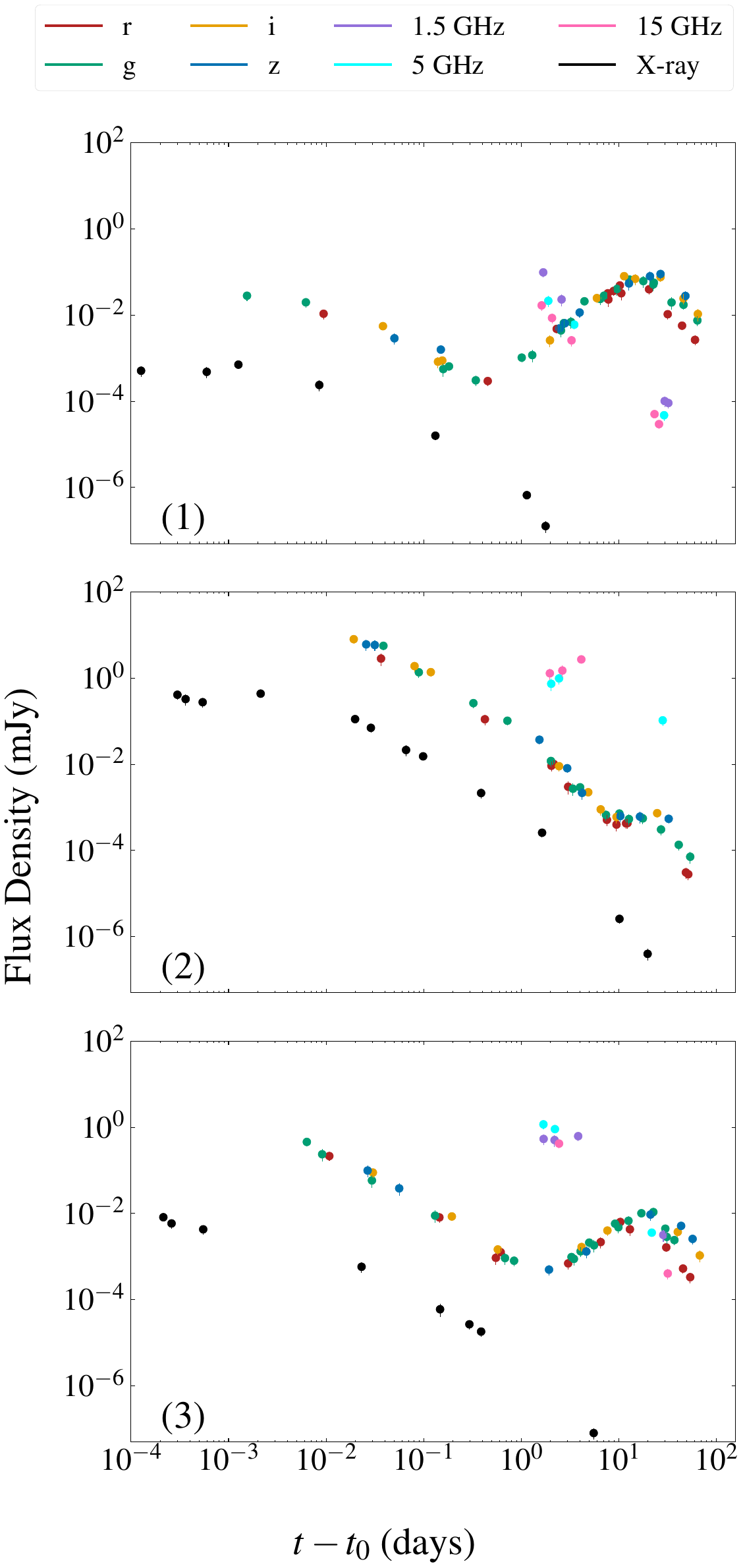}
\caption{Examples of simulated datasets with (1) a dominant SN, (2) a dominant GRB afterglow and (3) a typical GRB-SN event.}
\label{grbsn_example}
\end{figure}

From the 1000 simulated data sets, we saw three distinct morphological cases of GRB-SNe where (1) the SN was dominant over the GRB afterglow, (2) the GRB afterglow was dominant over the SN and (3) a stereotypical GRB-SN event where both components are clearly visible. Examples of each case can be found in Figure~\ref{grbsn_example}. Such cases were extracted from the 1000 simulated data sets on the basis of the ratio between the true SN flux and the true GRB flux around the peak of the SN. From the simulated parameters, we found that the median peak time in the $r$-band light curve was approximately 16.88 days. We then extracted the SN flux and the GRB flux at this time using the simulated parameters to produce a histogram of flux ratios. Simulated data sets that fall above the upper end of the credible interval of this distribution with a ratio of $\gtrsim 1 \times10^8$ produce the SN-dominated GRB-SNe, while the simulated datasets that fall below the credible interval with ratios less than $\lesssim 1.5$ afterglow dominated GRB-SNe. Simulated GRB-SNe that fall within the credible interval produce the stereotypical or normal GRB-SN light curves.

Another interesting feature of the simulated light curves was the variety of jet break times in the GRB afterglow. Some simulated light curves have clear jet breaks that occur early in the light curve when the GRB is dominant, while others occur later when the SN is dominant and is not observable in the optical bands. We believe that the positioning of the jet break could affect the modelling results. For example, a GRB-SN event that has an early jet break that is observable across optical, X-ray and radio bands could help better constrain the GRB parameters of the model than an event with a late jet break obscured in the optical bands by the SN emission.

Through the investigation of these three GRB-SN cases and the positioning of the jet break, we believe the joint modelling approach is the best in reproducing GRB-SN light curves. To demonstrate this, we compare our joint modelling approach to other fitting strategies. We perform the afterglow subtraction method by fitting all data bands with only the \texttt{tophat\_from\_emulator} model. We then subtract the best-fitting model from the simulated data to reveal the afterglow-subtracted SN and then fit this with the Arnett model. The subtracted data should include errors from both the original data and the errors introduced by the afterglow modelling. The errors in the \texttt{tophat\_from\_emulator} model can be estimated from the posteriors returned from the modelling. In this case, we find that the errors introduced by this modelling are much less than the simulated data error. As a result, we do not introduce an additional error term to the data to account for the uncertainty introduced by the subtraction of the afterglow to simplify our analysis. To evaluate this methodology alongside physically motivated alternatives, we fit the simulated datasets using: (a) the \texttt{tophat\_from\_emulator} model only (herein the tophat only model), (b) the Arnett model only, (c) the Arnett model after subtracting the best-fitting afterglow from the data, and (d) a joint fit in which the GRB afterglow and supernova models are fitted simultaneously. In the following sections,s we show an example of a fitted light curve for each case using each methodology. To directly compare the performance of each modelling methodology, we present violin plots showing the recovery of the injection parameters for twenty randomly selected events from each case (60 datasets in total). The distributions show the normalised difference between the posterior median and the injected value for each parameter. For the GRB parameters, the differences obtained from both the joint GRB-SN fits and the tophat only fits are normalised by the posterior standard deviation of the corresponding parameter from the tophat only fit. Similarly, for the SN parameters, the differences from the joint GRB-SN fits, afterglow-subtracted fits, and Arnett only fits are normalised by the posterior standard deviation obtained from the Arnett only fit.

\begin{figure*}[ht!]
\centering
\includegraphics[width=0.96\linewidth]{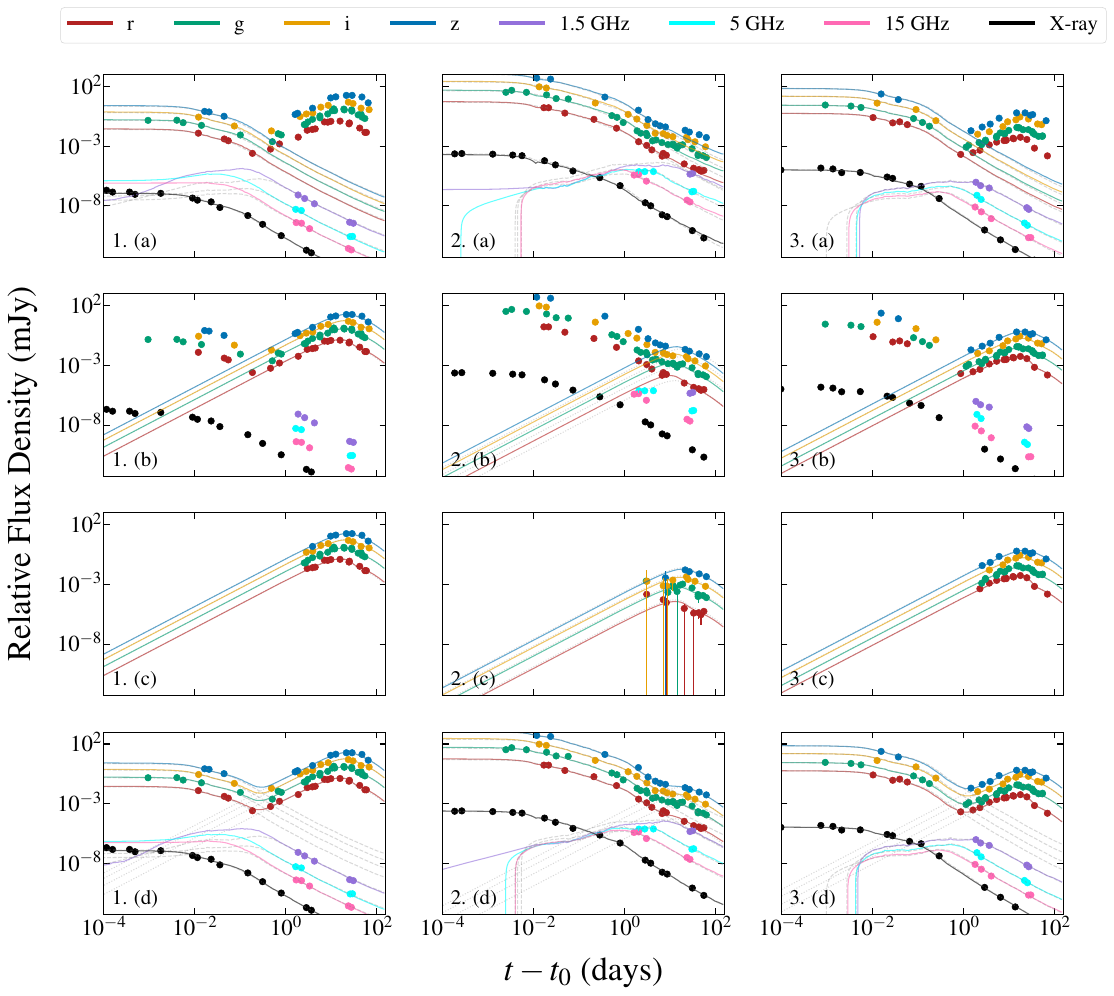}

\caption{Breakdown of fitting results for three chosen cases of GRB-SNe (columns) with early break times, using four different modelling approaches (rows). The columns are the (1) SN-dominated, (2) GRB-dominated, and (3) normal GRB-SN cases, respectively. The rows are the (a) tophat only model, (b) Arnett only model, (c) afterglow-subtracted method, and (d) our joint fitting model. The dashed grey lines represent the true GRB light curve and the dotted grey lines represent the true SN light curve produced using the injection parameters. Each band is vertically offset for clarity.}
\label{fitting_all_earlytb}

\end{figure*}

\subsection{Case 1: Early jet break}\label{case1}

In this case, we compare how each modelling strategy behaves with the three GRB-SN scenarios, each with an early jet break. Figure~\ref{fitting_all_earlytb} shows that the tophat only, Arnett only, and afterglow subtractions all seem to work well in fitting the light curve for the SN-dominated and normal GRB-SN cases shown in the first and third columns. When it comes to the afterglow-dominated GRB-SN event in the second column, however, the tophat only model fit (2.(a)) shows that the model overpredicts the afterglow at late times, while the Arnett only model (2.(b)) predicts that the SN peak occurs earlier and brighter than the true SN light curve. The errors in the afterglow model then translate across to the afterglow-subtracted model (2.(c)) where the SN contribution is now underpredicted. In all three cases, the joint fit performs well in reproducing the true behaviour of each component, as seen in the bottom row of Figure~\ref{fitting_all_earlytb}.

\begin{figure*}[h!]
\centering
\includegraphics[width=0.82\linewidth]{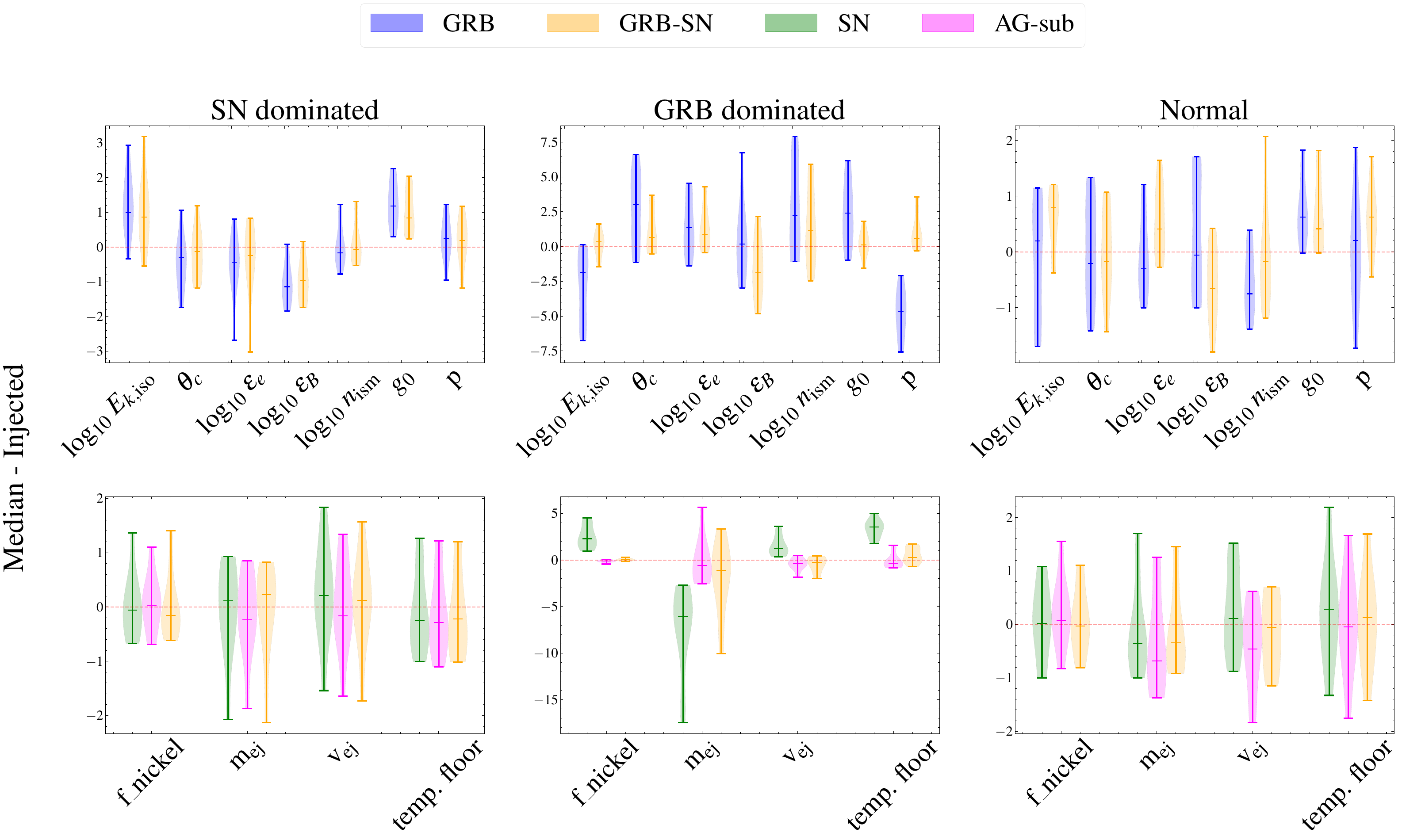}
\caption{Violin plots representing the performance of each modelling method in predicting the injected parameter for twenty simulated datasets for each of the three GRB-SN cases with an early jet break.}
\label{all_vp_early}
\end{figure*}

The violin plots in Figure~\ref{all_vp_early} show that all models describe the data well for the SN-dominated case, with the joint fit performing slightly better than the tophat only model when predicting the GRB parameters. Both models fail to return the injection value for the Lorentz factor $g_0$ in the SN-dominated case, which could be due to a lack of pure afterglow data at early times. For the GRB-dominated case, the joint fit is much better at reproducing the injection parameters, as seen in the middle column of Figure~\ref{all_vp_early}. The tophat only model completely fails to reproduce the $p$ parameter, while the Arnett only model fails to reproduce all of the injection parameters. All models succeed in returning most of the GRB and SN parameters in the case of the normal GRB-SN in the final column of Figure~\ref{all_vp_early}; however, the models still struggle slightly in returning the $g_0$ parameter. In all cases, the afterglow subtraction method can reproduce the SN parameters, but again, caution must be taken when interpreting this, as it inherits error from the modelling of the GRB afterglow that is not taken into account in the modelling.

\begin{figure*}[h!]
\centering
\includegraphics[width=0.82\linewidth]{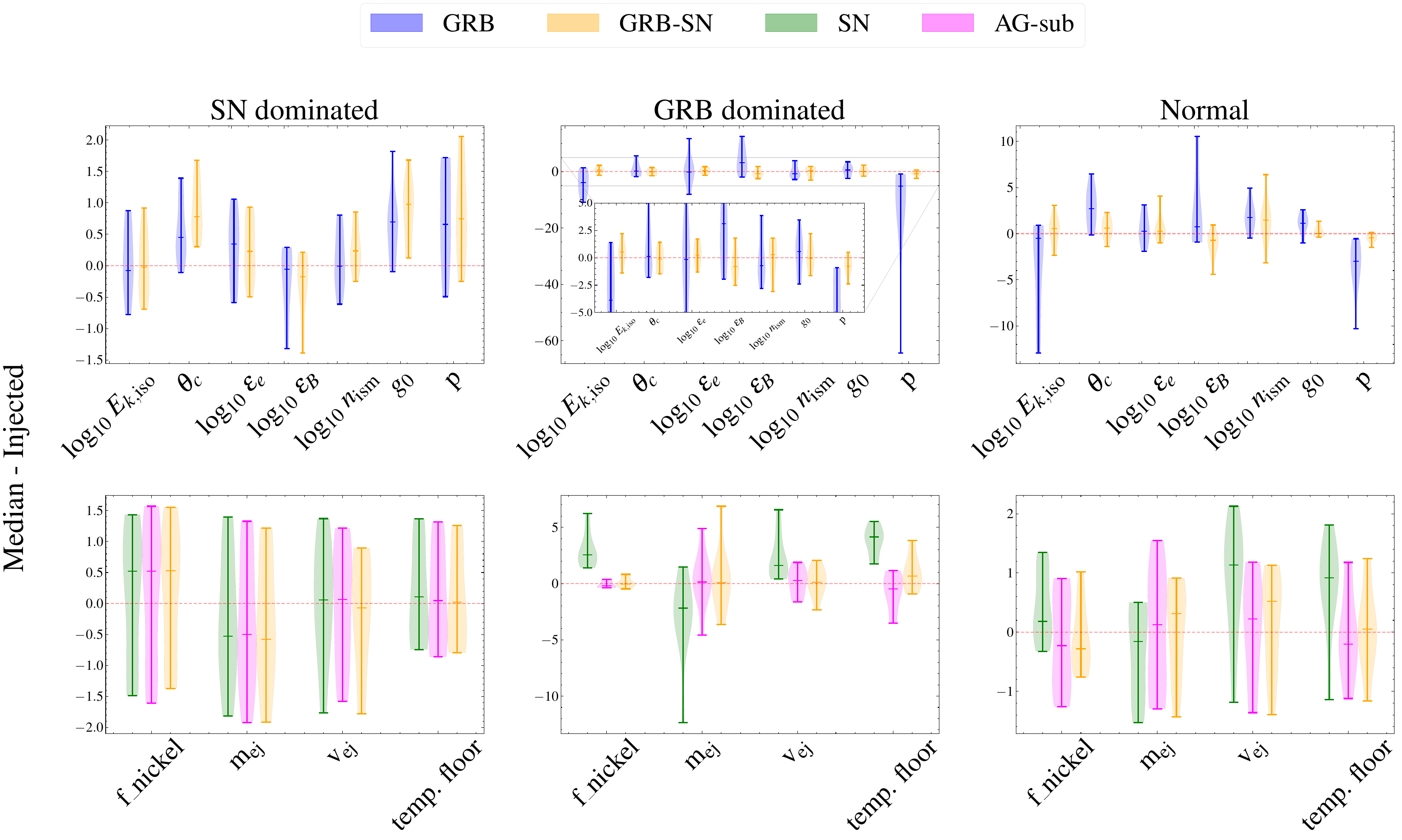}
\caption{Violin plots representing the performance of each modelling method in predicting the injected parameter for twenty simulated datasets for each of the three GRB-SN cases with a late jet break.}
\label{all_vp_late}
\end{figure*}

\subsection{Case 2: Late break time}\label{case2}

We repeat the same methodology for cases with a late jet break that occurs when the SN is dominant and is thus obscured in the optical bands. We see results similar to the early jet-break case, where the SN-dominated and the normal GRB-SN cases match the true fits well. Again, we see noticeable deviations from the true GRB afterglow and SN behaviour in the GRB-dominated case. The joint fit reproduced the light curve well for all three cases. The plot for the late jet break time case can be found in the appendix in Figure~\ref{fitting_all_latetb}.

Even though the models are observed to fit well, the violin plots across the subsamples reveal much more information about how the modelling is behaving. In contrast to the SN-dominated case with an early jet break, Figure~\ref{all_vp_late} shows that both modelling methods struggle to return both the $\theta_c$ and $g_0$ parameters. When considering the GRB-dominated case with a late jet break shown in Figure~\ref{all_vp_early}, the results are similar to the early jet break case, where the Arnett only and tophat only models fail to reproduce some of the injected parameters. The Arnett only model performs slightly better than the early break time case shown in the second column of Figure~\ref{all_vp_early}, where it more accurately predicts the $M_\textbf{ej}$ parameter; however, it fails to return the other parameters. For the normal GRB-SN case, each modelling methodology reproduces most of the GRB and SN parameters well; however, the tophat only model fails to return the true $p$ value. Interestingly, we see a deviation in the behaviour of the joint model from the early jet break case, seen in the third column of Figure~\ref {all_vp_early}, where the joint model can predict the $g_0$ parameter very well.

\begin{figure*}[h!]
\centering
\includegraphics[width=0.97\linewidth]{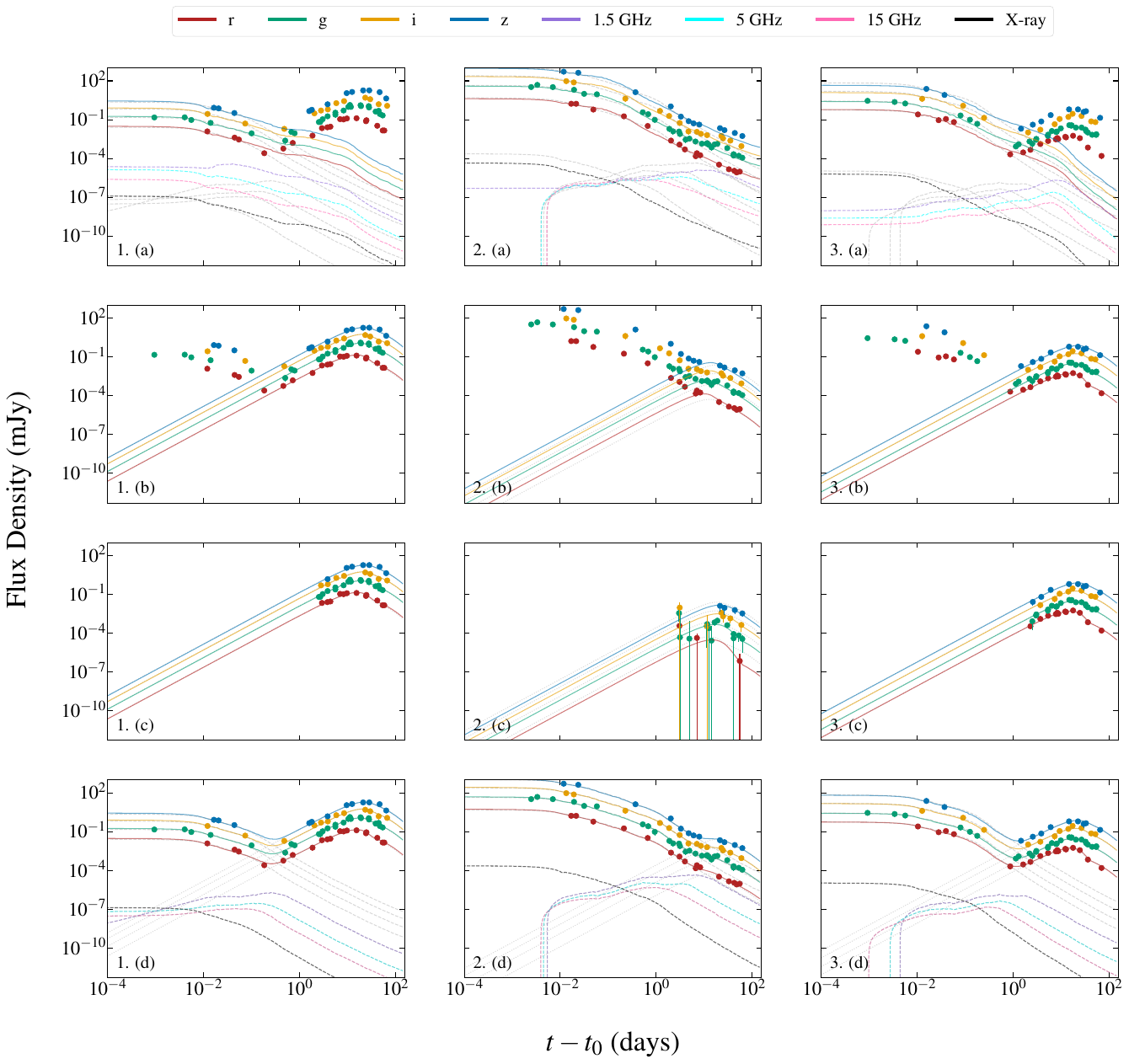}

\caption{Breakdown of fitting results for three chosen cases of GRB-SNe (columns) with early jet break times, using four different modelling approaches (rows) where no X-ray or radio data were used in the modelling. The columns are the (1) SN-dominated, (2) GRB-dominated, and (3) normal GRB-SN cases, respectively. The rows are the (a) tophat only model, (b) Arnett only model, (c) afterglow-subtracted method, and (d) our joint fitting model. The dashed grey lines represent the true GRB light curve and the dotted grey lines represent the true SN light curve produced using the injection parameters. Each band is vertically offset for clarity.}
\label{Fitting_no_Xray_earlytb}

\end{figure*}

\begin{figure*}[h!]
\centering
\includegraphics[width=0.82\linewidth]{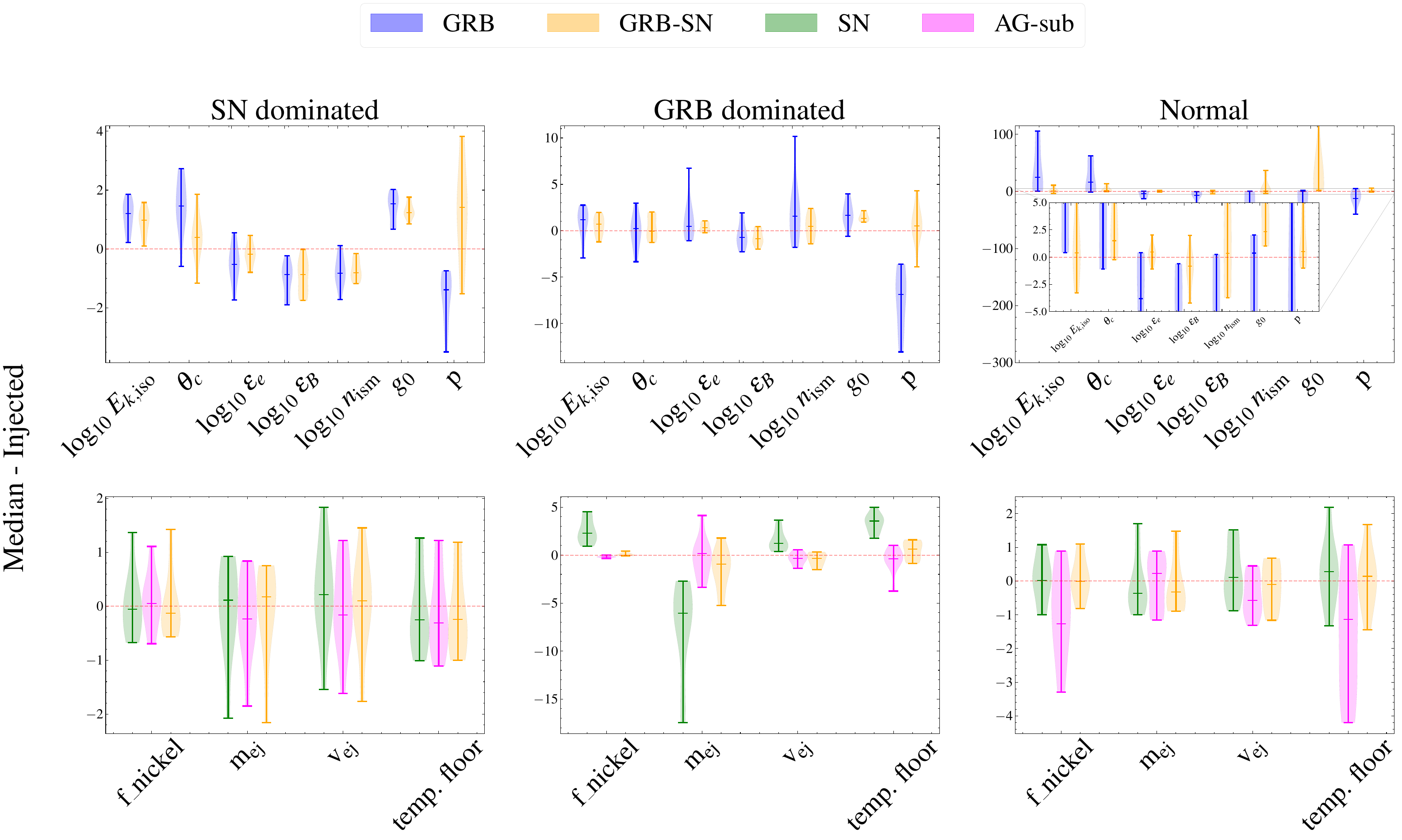}

\caption{Breakdown of fitting results for three chosen cases of GRB-SNe with an early break with no X-ray and Radio data included in the fitting.}
\label{all_vp_plots_early_tb_noXray_noRadio}

\end{figure*}

\begin{figure*}[h!]
\centering
\includegraphics[width=0.82\linewidth]{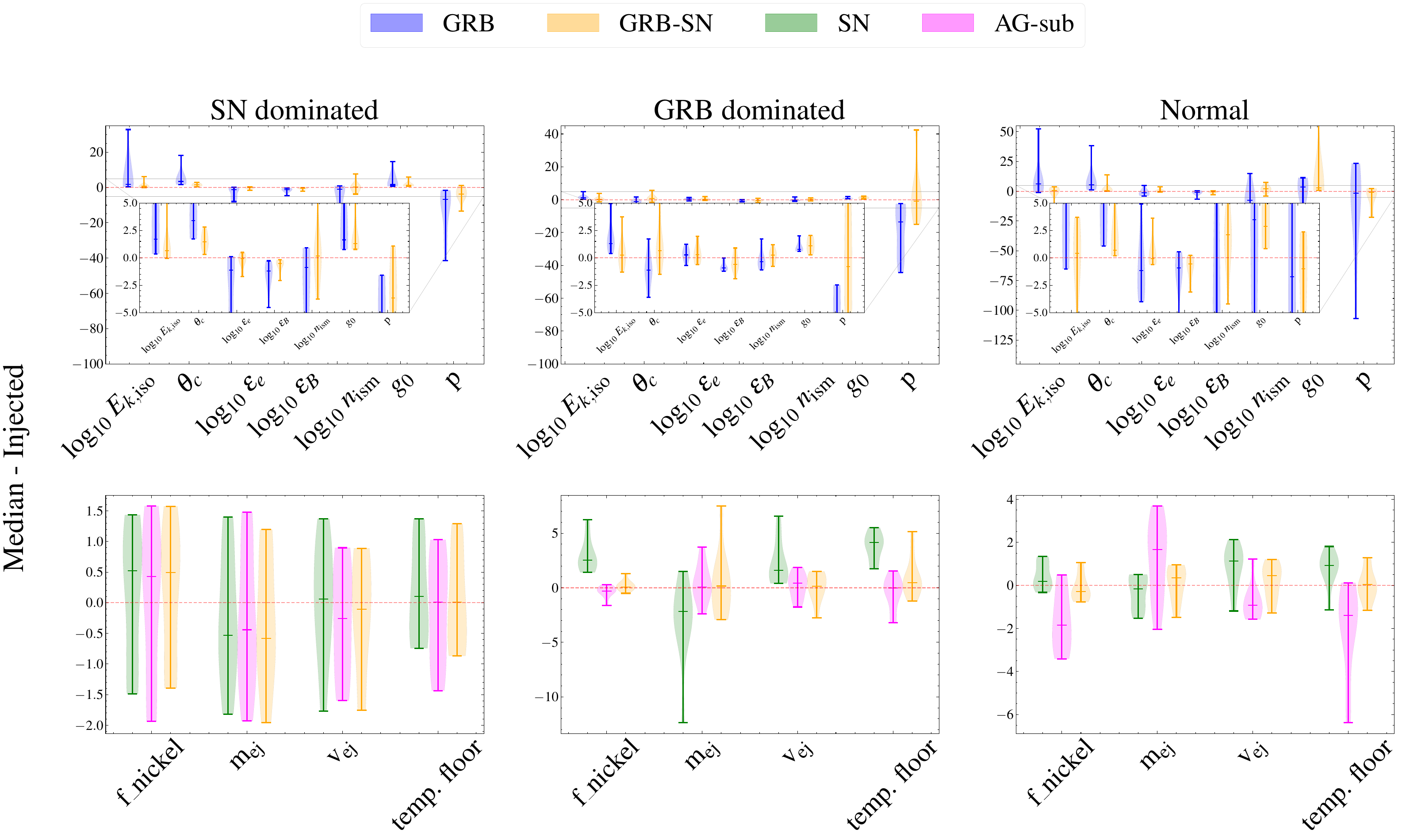}

\caption{Breakdown of fitting results for three chosen cases of GRB-SNe with a late break with no X-ray and Radio data included in the fitting.}
\label{all_vp_plots_late_tb_noXray_noRadio}

\end{figure*}

\subsubsection{Case 3: Fitting with no X-ray or Radio}

In the previous case, each GRB-SN data point is well populated with data in the optical, X-ray and radio bands. In this case, we investigate the model behaviour when X-ray and radio bands are not available, to explore the importance of these bands in constraining parameter values. The best-fit parameters obtained from fitting only the optical bands are used to produce synthetic X-ray and radio light curves, represented by the coloured dashed lines. The same example data sets are used in the modelling as in the previous sections.

The tophat only model struggles when the X-ray and radio bands are removed, overpredicting the afterglow contribution at later times as seen in the first row of Figure~\ref{Fitting_no_Xray_earlytb}. The Arnett only and afterglow-subtracted model fitting still seems to reproduce the light curves well, as seen in Figure~\ref{fitting_all_earlytb} where the X-ray and radio bands were included in the model.  Similar to cases where X-ray and radio bands were included in the modelling, the tophat only, Arnett only, and afterglow-subtracted models struggle to fit the GRB-dominated light curve, as shown in the second column of Figure~\ref{Fitting_no_Xray_earlytb}.

The violin plots in Figure~\ref{all_vp_plots_early_tb_noXray_noRadio} show that for each case the $g_0$ parameter cannot be accurately predicted by either the joint or the tophat only model when X-ray and radio bands are excluded from the modelling. The results of the SN-dominated case suffer the most when X-ray and radio bands are excluded with bith the joint and tophat only model failing to reproduce the $\text{log}{10}\, E_{k,\text{iso}}$, $\text{log}{10}\,\epsilon_B$, $\text{log}{10}\,n_\text{ism}$ and $g_0$ parameters. For the GRB-dominated and the normal GRB-SN cases, the joint model performs much better than the tophat only model, returning all parameter values except $g_0$. The results of the SN modelling for each case remain similar to those obtained in the previous cases where X-ray and radio bands were included, with parameters being returned successfully in all cases bar the GRB-dominated case.

Again, we see that the modelling results suffer when X-ray and radio bands are not included in the modelling of GRB-SNe with late jet break times. The fits for the late jet break scenario show the same behaviour as the early jet break scenario where, again, the tophat only model struggles to reproduce the expected fit across all three cases. The corresponding late jet break fits are presented in Appendix Figure~\ref{Fitting_no_Xray_latetb}.

Just like the early jet break scenario, Figure~\ref{all_vp_plots_late_tb_noXray_noRadio} shows that the $g_0$ parameter is not successfully recovered in any of the three cases when using either the joint or tophat only models. Interestingly, both the joint and tophat only models struggle to reproduce the $\theta_c$ parameter in the SN-dominated and normal GRB-SN cases, which is not seen in the previous early jet break scenario. The remaining GRB and SN parameters show broadly similar behaviour to that found for the early jet break scenario.

\begin{figure*}[h!]
\centering

\begin{subfigure}{0.32\linewidth}
    \centering
    \includegraphics[width=\linewidth]{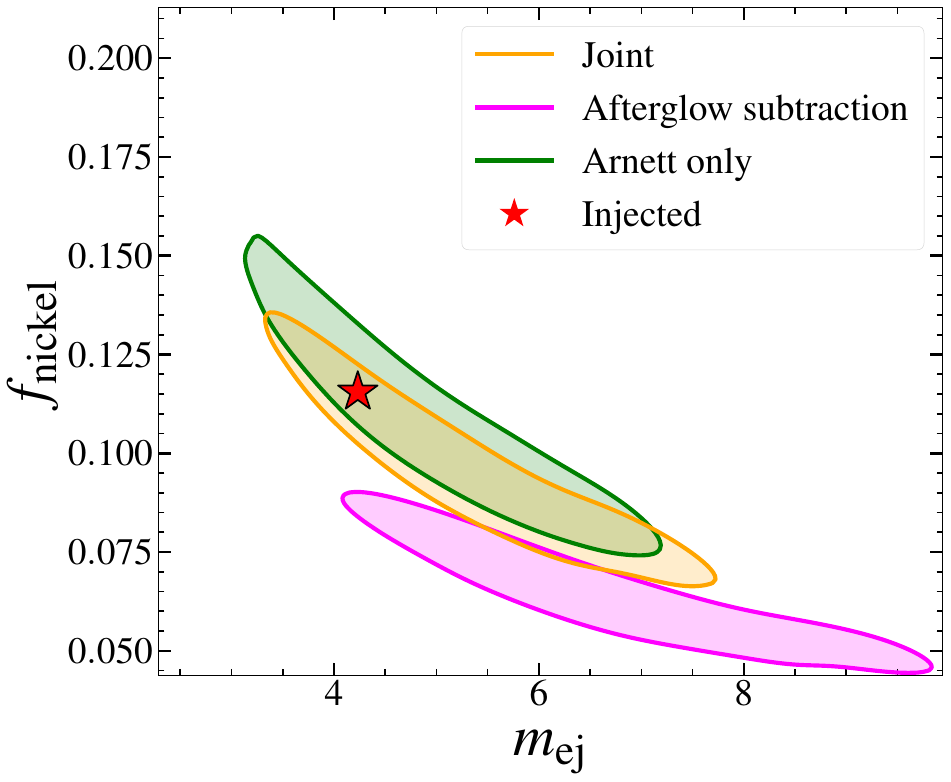}
    \label{fnimej}
\end{subfigure}
\hfill
\begin{subfigure}{0.32\linewidth}
    \centering
    \includegraphics[width=\linewidth]{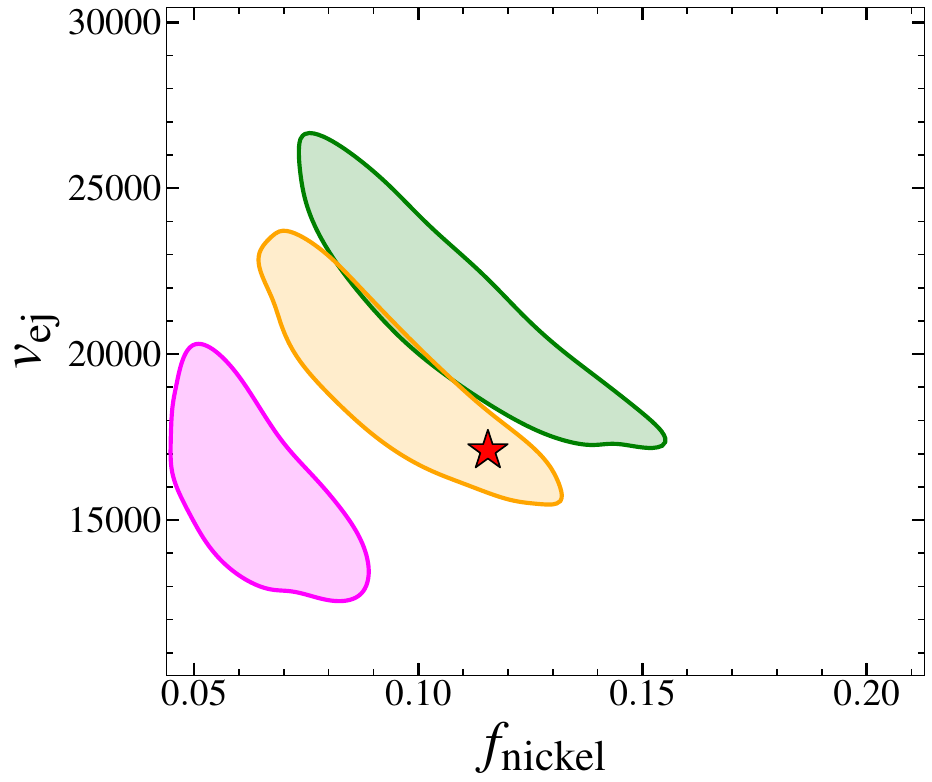}
    \label{fnivej}
\end{subfigure}
\hfill
\begin{subfigure}{0.32\linewidth}
    \centering
    \includegraphics[width=\linewidth]{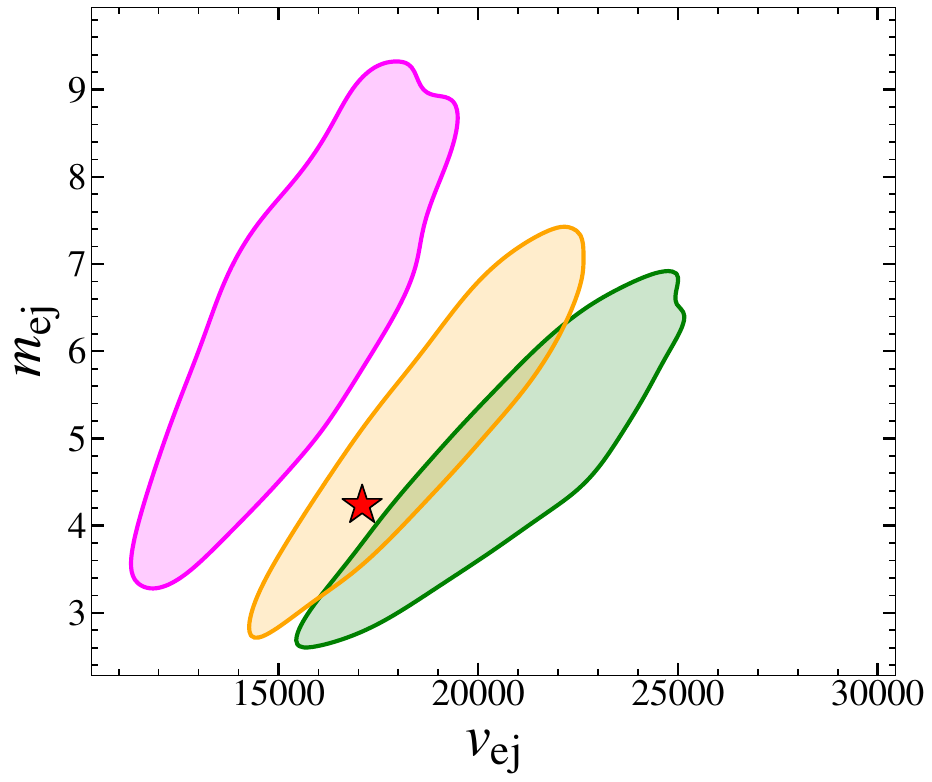}
    \label{vejmej}
\end{subfigure}

\caption{Contour plots displaying correlations between the SN parameters obtained from the multi-band fitting of the normal GRB-SN case with a late jet break with each modelling strategy. Each region represents the $1\sigma$ credible interval of the posterior.}
\label{corner_SN}
\end{figure*}

\section{Discussion and Conclusions}

In this work, we have explored four fitting strategies used to describe GRB-SN events in an attempt to reproduce the true GRB and SN injection parameters. We focused on events with early and late jet break times for three different cases with varying GRB/SN contamination: (1) where the SN dominates, (2) where the GRB dominates and (3) a normal GRB-SN where the contributions are relatively balanced. The four different modelling methodologies employed in this work were fitting data for each case with (a) a tophat only model, (b) an Arnett only model, (c) an afterglow-subtracted model and (d) a joint tophat and Arnett model.

Although the tophat only and Arnett only models frequently produced visually acceptable fits, they did not always recover the true injected parameters. This limitation was most pronounced in the GRB-dominated case in both the late and early jet break scenarios, where the bright afterglow masks the emergence of the SN and obscures key light-curve features such as the jet break and SN peak. We observed in both the early and late jet break cases that even though the results of the tophat only and Arnett only models produce good fits to the data, the distribution parameters returned by these models can sometimes fail to reproduce the true injection parameters of the data. The joint fit model reproduces both GRB and SN parameters efficiently, whereas the tophat only and Arnett only models fail to do so. Both the early and late jet break cases exhibited broadly similar behaviour, with the main exception that the tophat only model struggled to recover the electron energy distribution index, $p$, for the normal GRB-SN case with a late jet break. 

It is important to note that the joint modelling struggles to reproduce the $g_0$ and sometimes the $\theta_c$ parameters. Both $g_0$ and $\theta_c$ are strongly constrained by the location of and the concentration of data before and after the jet break in the GRB afterglow. For example, in the SN-dominated case, there is little information on the afterglow as the SN emission dominates the optical bands. In contrast, it is much easier to get an accurate handle on the afterglow behaviour in the GRB-dominated case. The $\theta_c$ parameter is not returned accurately by the joint model in the SN-dominated case of the late jet break. We interpret this as a consequence of the limited temporal coverage around the jet break and of the SN component obscuring the jet-break signature in the optical bands. The jet break location and flux are sensitive to parameters such as $E_\text{k,iso}$, $n_\text{ism}$, $\epsilon_e$ and $\epsilon_B$ \citep{1999_Sari,Wu_2004,2009_racusin}. In the absence of sufficient early-time and broadband data, multiple combinations of these parameters can produce similar break times and fluxes, leading to an incorrect estimate or a broad posterior for $g_0$ or $\theta_c$. This degeneracy is particularly important when the optical light curve is contaminated by an underlying SN. Previous modelling studies likewise show that uncontaminated radio and X-ray observations substantially improve afterglow-parameter recovery, whereas ignoring these bands broadens the posterior distributions and can increase parameter biases \citep{Wallace_2025}. Improved X-ray and radio cadence around the expected jet-break epoch would therefore help constrain the afterglow independently of the SN component.

In our simulations, when the joint model fails to recover $g_0$, the parameter is consistently overpredicted. This systematic behaviour in the joint modelling provides a useful caveat when interpreting $g_0$: although the inferred value should not be considered a reliable estimate of the true initial Lorentz factor, the consistent direction of the bias indicates that the true value is likely to be lower than that is returned by the model. Thus, users can interpret the inferred $g_0$ with appropriate caution, recognising that it may represent an upper-biased estimate of the initial Lorentz factor. This may simply be pointing towards a failure of the underlying emulator~\citep{Wallace_2025}, which is an approximation to the full numerical model~\citep{lamb_2018}.

We also found that the inclusion of X-ray and radio bands is constraining towards GRB parameters when performing multi-band modelling.  Without these additional wavelength constraints, the tophat only model overpredicts the GRB contribution in all cases for both early and late jet break scenarios, as the model could no longer reliably distinguish between the GRB and SN emission. While the joint model continues to provide good fits to the observed light curves, the violin plots show that it is unable to reliably recover the $g_0$ parameter across any of the GRB-SN cases. The parameter recovery is poorest for the SN-dominated case, where the joint model fails to accurately recover several of the injected parameters for both the early and late jet break scenarios. In the late jet break scenario, the $\theta_c$ parameter is particularly poorly constrained. We believe that this degradation in the joint modelling is primarily caused by contamination from the SN component, which makes determining the location and characteristics of the jet break considerably more difficult when only optical observations are available. The joint model could produce an accurate parameter estimation in cases with an early jet break; however, this would require high-cadence observations within the first day after the burst, before the SN begins to dominate. Obtaining such observations is challenging, particularly because early X-ray localisation and follow-up observations are generally required to identify and characterise the afterglow during this period. We thus conclude that X-ray and radio data are vital to accurately characterise the relativistic jet and SN parameters.

In all cases, we included the strategy of subtracting the afterglow from the light curve to reveal the underlying SN. While this approach often recovered the SN parameters successfully, it is fundamentally inconsistent from a Bayesian perspective. This method assumes that the afterglow model is the true afterglow, treating the best-fitting model as exact rather than one realisation drawn from a posterior distribution. As a result, the uncertainty from the afterglow modelling is not propagated into the subtraction. This causes the resulting light curve to inherit the uncertainties from the afterglow fit, and the subsequent SN modelling treats this data as though it were measured independently. This is not representative of how the true GRB-SN data are generated and thus constitutes a misspecified model. The bias inherited by the afterglow subtraction method is exemplified by the contour plots in Figure~\ref{corner_SN}. In this figure, we show the posterior distributions obtained from the complete multi-band fitting for the normal GRB-SN case with a late jet break using each modelling strategy. For each parameter, the joint model can correctly predict the true injection parameters within the $1\sigma$ credible interval, whereas the afterglow subtraction and Arnett only modelling posteriors are biased away from the true injection parameters when considering this same interval. This emphasises that the joint modelling approach provides a much more robust and accurate prediction of the true injection parameters. Unlike the afterglow subtraction method, the joint modelling propagates the uncertainties for both the GRB afterglow and the SN components, providing a statistically consistent parameter estimate and a more rigorous Bayesian treatment of GRB-SN observations.

Overall, this study demonstrates that robust parameter inference for GRB-SNe and, by extension, FXT-SNe, requires simultaneous modelling of both the jet and SN components. Although independent jet or SN models can often reproduce the observed light curves, they do not necessarily recover the underlying physical parameters, particularly when one component contaminates the other. In addition to this, some GRB/FXT-SNe may have additional thermal components present in the early parts of their light curves, which can further affect the estimation of the afterglow parameters \citep[GRB\,060218/SN\,2006aj, EP250304a/SN\,2025fhm;][]{Campana_2006,cotter2026n}. Joint modelling naturally accounts for the contribution and uncertainty of each initial component, allowing the full information contained within the multi-wavelength dataset to be exploited while avoiding the biases introduced by treating individual components in isolation. Our results emphasise the importance of broad wavelength coverage. X-ray and radio observations provide important constraints on the jet evolution and break degeneracies that cannot be resolved using optical data alone. Without these bands, the jet and SN components become increasingly difficult to disentangle, leading to poor constraints on the jet parameters, particularly for SN-dominated events. These findings reinforce the need for coordinated multi-wavelength follow-up campaigns that obtain X-ray, optical, and radio observations from the earliest possible epochs.
As the number of well-observed FXT/GRB-SNe continues to grow through facilities such as the \textit{EP} mission and the Vera C. Rubin Observatory, statistically rigorous joint modelling frameworks will become increasingly important. Such approaches will enable more reliable measurements of explosion properties, ultimately leading to a deeper understanding of the connection between relativistic jets, core-collapse SNe, and their common central engine.

\begin{acknowledgements}
LC and AMC acknowledge the support of the Irish Research Council Postgraduate Scholarship No.
GOIPG/2022/1008. NS acknowledges support from the Kavli Foundation. 
\end{acknowledgements}

\bibliographystyle{aa} % style aa.bst
\bibliography{ref}

@ARTICLE{woosley2006supernova,
  author = {Woosley, S. E. and Bloom, J. S.},
  title = {The supernova--gamma-ray burst connection},
  journal = {Annual Review of Astronomy and Astrophysics},
  volume = {44},
  pages = {507--556},
  year = {2006}
}

@book{hjorth2012grb,
  title={The GRB-supernova connection},
  author={Hjorth, Jens and Bloom, Joshua S},
  volume={51},
  year={2012},
  publisher={Cambridge University Press Cambridge; New York}
}

@article{cano2017observer,
  title={The Observer’s Guide to the Gamma-Ray Burst Supernova Connection},
  author={Cano, Zach and Wang, Shan-Qin and Dai, Zi-Gao and Wu, Xue-Feng},
  journal={Advances in Astronomy},
  volume={2017},
  number={1},
  pages={8929054},
  year={2017},
  publisher={Wiley Online Library}
}

@ARTICLE{Sarin2021,
       author = {{Sarin}, Nikhil and {Ashton}, Gregory and {Lasky}, Paul D. and {Ackley}, Kendall and et al.},
        title = "{CDF-S XT1: The off-axis afterglow of a neutron star merger at $z=2.23$}",
      journal = {arXiv e-prints},
         year = 2021,
        month = may,
          eid = {arXiv:2105.10108},
        pages = {arXiv:2105.10108},
          doi = {10.48550/arXiv.2105.10108},
archivePrefix = {arXiv},
       eprint = {2105.10108},
 primaryClass = {astro-ph.HE},
       adsurl = {https://ui.adsabs.harvard.edu/abs/2021arXiv210510108S}
}

@ARTICLE{Gal1998bw,
       author = {{Galama}, T.~J. and {Vreeswijk}, P.~M. and {van Paradijs}, J. and {Kouveliotou}, C. and {Augusteijn}, T. and {B{\"o}hnhardt}, H. and {Brewer}, J.~P. and {Doublier}, V. and {Gonzalez}, J. -F. and {Leibundgut}, B. and {Lidman}, C. and {Hainaut}, O.~R. and {Patat}, F. and {Heise}, J. and {in't Zand}, J. and {Hurley}, K. and {Groot}, P.~J. and {Strom}, R.~G. and {Mazzali}, P.~A. and {Iwamoto}, K. and {Nomoto}, K. and {Umeda}, H. and {Nakamura}, T. and {Young}, T.~R. and {Suzuki}, T. and {Shigeyama}, T. and {Koshut}, T. and {Kippen}, M. and {Robinson}, C. and {de Wildt}, P. and {Wijers}, R.~A.~M.~J. and {Tanvir}, N. and {Greiner}, J. and {Pian}, E. and {Palazzi}, E. and {Frontera}, F. and {Masetti}, N. and {Nicastro}, L. and {Feroci}, M. and {Costa}, E. and {Piro}, L. and {Peterson}, B.~A. and {Tinney}, C. and {Boyle}, B. and {Cannon}, R. and {Stathakis}, R. and {Sadler}, E. and {Begam}, M.~C. and {Ianna}, P.},
        title = "{An unusual supernova in the error box of the {\ensuremath{\gamma}}-ray burst of 25 April 1998}",
      journal = {\nat},
         year = 1998,
        month = oct,
       volume = {395},
       number = {6703},
        pages = {670-672},
          doi = {10.1038/27150},
archivePrefix = {arXiv},
       eprint = {astro-ph/9806175},
 primaryClass = {astro-ph},
       adsurl = {https://ui.adsabs.harvard.edu/abs/1998Natur.395..670G}
}

@article{finneran_2025_webtool,
title = {The GRBSN webtool: An open-source repository for gamma-ray burst-supernova associations},
journal = {Astronomy and Computing},
volume = {52},
pages = {100954},
year = {2025},
issn = {2213-1337},
doi = {https://doi.org/10.1016/j.ascom.2025.100954},
url = {https://www.sciencedirect.com/science/article/pii/S2213133725000277},
author = {Gabriel Finneran and Laura Cotter and Antonio Martin-Carrillo}

}

@ARTICLE{1998_sari_piran,
       author = {{Sari}, Re'em and {Piran}, Tsvi and {Narayan}, Ramesh},
        title = "{Spectra and Light Curves of Gamma-Ray Burst Afterglows}",
      journal = {\apjl},
         year = 1998,
        month = apr,
       volume = {497},
       number = {1},
        pages = {L17-L20},
          doi = {10.1086/311269},
archivePrefix = {arXiv},
       eprint = {astro-ph/9712005},
 primaryClass = {astro-ph},
       adsurl = {https://ui.adsabs.harvard.edu/abs/1998ApJ...497L..17S}
}

@article{arnett1982,
  title={Type I supernovae. I-Analytic solutions for the early part of the light curve},
  author={Arnett, W David},
  journal={Astrophysical Journal, Part 1, vol. 253, Feb. 15, 1982, p. 785-797.},
  volume={253},
  pages={785--797},
  year={1982}
}

@ARTICLE{Redback,
       author = {{Sarin}, Nikhil and {H{\"u}bner}, Moritz and {Omand}, Conor M.~B. and {Setzer}, Christian N. and {Schulze}, Steve and {Adhikari}, Naresh and {Sagu{\'e}s-Carracedo}, Ana and {Galaudage}, Shanika and {Wallace}, Wendy F. and {Lamb}, Gavin P. and {Lin}, En-Tzu},
        title = "{REDBACK: a Bayesian inference software package for electromagnetic transients}",
      journal = {\mnras},
         year = 2024,
        month = jun,
       volume = {531},
       number = {1},
        pages = {1203-1227},
          doi = {10.1093/mnras/stae1238},
archivePrefix = {arXiv},
       eprint = {2308.12806},
 primaryClass = {astro-ph.HE},
       adsurl = {https://ui.adsabs.harvard.edu/abs/2024MNRAS.531.1203S}
}

@ARTICLE{bilby,
       author = {{Ashton}, Gregory and {H{\"u}bner}, Moritz and {Lasky}, Paul D. and {Talbot}, Colm and {Ackley}, Kendall and {Biscoveanu}, Sylvia and {Chu}, Qi and {Divakarla}, Atul and {Easter}, Paul J. and {Goncharov}, Boris and {Hernandez Vivanco}, Francisco and {Harms}, Jan and {Lower}, Marcus E. and {Meadors}, Grant D. and {Melchor}, Denyz and {Payne}, Ethan and {Pitkin}, Matthew D. and {Powell}, Jade and {Sarin}, Nikhil and {Smith}, Rory J.~E. and {Thrane}, Eric},
        title = "{BILBY: A User-friendly Bayesian Inference Library for Gravitational-wave Astronomy}",
      journal = {\apjs},
         year = 2019,
        month = apr,
       volume = {241},
       number = {2},
          eid = {27},
        pages = {27},
          doi = {10.3847/1538-4365/ab06fc},
archivePrefix = {arXiv},
       eprint = {1811.02042},
 primaryClass = {astro-ph.IM},
       adsurl = {https://ui.adsabs.harvard.edu/abs/2019ApJS..241...27A}
}

@article{Campana_2006,
    author = "Campana, S. and others",
    title = "{The shock break-out of grb 060218/sn 2006aj}",
    eprint = "astro-ph/0603279",
    archivePrefix = "arXiv",
    doi = "10.1038/nature04892",
    journal = "Nature",
    volume = "442",
    pages = "1008--1010",
    year = "2006"
}

@ARTICLE{lamb_2018,
       author = {{Lamb}, Gavin P. and {Mandel}, Ilya and {Resmi}, Lekshmi},
        title = "{Late-time evolution of afterglows from off-axis neutron star mergers}",
      journal = {\mnras},
         year = 2018,
        month = dec,
       volume = {481},
       number = {2},
        pages = {2581-2589},
          doi = {10.1093/mnras/sty2196},
archivePrefix = {arXiv},
       eprint = {1806.03843},
 primaryClass = {astro-ph.HE},
       adsurl = {https://ui.adsabs.harvard.edu/abs/2018MNRAS.481.2581L}
}

@ARTICLE{Taddia_2019,
       author = {{Taddia}, F. and {Sollerman}, J. and {Fremling}, C. and {Barbarino}, C. and {Karamehmetoglu}, E. and {Arcavi}, I. and {Cenko}, S.~B. and {Filippenko}, A.~V. and {Gal-Yam}, A. and {Hiramatsu}, D. and {Hosseinzadeh}, G. and {Howell}, D.~A. and {Kulkarni}, S.~R. and {Laher}, R. and {Lunnan}, R. and {Masci}, F. and {Nugent}, P.~E. and {Nyholm}, A. and {Perley}, D.~A. and {Quimby}, R. and {Silverman}, J.~M.},
        title = "{Analysis of broad-lined Type Ic supernovae from the (intermediate) Palomar Transient Factory}",
      journal = {\aap},
         year = 2019,
        month = jan,
       volume = {621},
          eid = {A71},
        pages = {A71},
          doi = {10.1051/0004-6361/201834429},
archivePrefix = {arXiv},
       eprint = {1811.09544},
 primaryClass = {astro-ph.HE},
       adsurl = {https://ui.adsabs.harvard.edu/abs/2019A&A...621A..71T}
}

@ARTICLE{Wallace_2025,
       author = {{Wallace}, Wendy F. and {Sarin}, Nikhil},
        title = "{A detailed dive into fitting strategies for GRB afterglows with contamination: a case study with kilonovae}",
      journal = {\mnras},
         year = 2025,
        month = jun,
       volume = {539},
       number = {4},
        pages = {3319-3335},
          doi = {10.1093/mnras/staf623},
archivePrefix = {arXiv},
       eprint = {2409.07539},
 primaryClass = {astro-ph.HE},
       adsurl = {https://ui.adsabs.harvard.edu/abs/2025MNRAS.539.3319W}
}

@ARTICLE{ashall2019,
       author = {{Ashall}, C. and {Mazzali}, P.~A. and {Pian}, E. and {Woosley}, S.~E. and {Palazzi}, E. and {Prentice}, S.~J. and {Kobayashi}, S. and {Holmbo}, S. and {Levan}, A. and {Perley}, D. and {Stritzinger}, M.~D. and {Bufano}, F. and {Filippenko}, A.~V. and {Melandri}, A. and {Oates}, S. and {Rossi}, A. and {Selsing}, J. and {Zheng}, W. and {Castro-Tirado}, A.~J. and {Chincarini}, G. and {D'Avanzo}, P. and {De Pasquale}, M. and {Emery}, S. and {Fruchter}, A.~S. and {Hurley}, K. and {Moller}, P. and {Nomoto}, K. and {Tanaka}, M. and {Valeev}, A.~F.},
        title = "{GRB 161219B/SN 2016jca: a powerful stellar collapse}",
      journal = {\mnras},
         year = 2019,
        month = aug,
       volume = {487},
       number = {4},
        pages = {5824-5839},
          doi = {10.1093/mnras/stz1588},
archivePrefix = {arXiv},
       eprint = {1702.04339},
 primaryClass = {astro-ph.HE},
       adsurl = {https://ui.adsabs.harvard.edu/abs/2019MNRAS.487.5824A}
}

@article{2009_racusin,
       author = {{Racusin}, J.~L. and {Liang}, E.~W. and {Burrows}, D.~N. and {Falcone}, A. and {Sakamoto}, T. and {Zhang}, B.~B. and {Zhang}, B. and {Evans}, P. and {Osborne}, J.},
        title = "{Jet Breaks and Energetics of Swift Gamma-Ray Burst X-Ray Afterglows}",
      journal = {\apj},
         year = 2009,
        month = jun,
       volume = {698},
       number = {1},
        pages = {43-74},
          doi = {10.1088/0004-637X/698/1/43},
archivePrefix = {arXiv},
       eprint = {0812.4780},
 primaryClass = {astro-ph},
       adsurl = {https://ui.adsabs.harvard.edu/abs/2009ApJ...698...43R}
}

@ARTICLE{2007_Medved,
       author = {{Medvedev}, Mikhail V.},
        title = "{Electron acceleration in relativistic GRB shocks}",
      journal = {Philosophical Transactions of the Royal Society of London Series A},
         year = 2007,
        month = may,
       volume = {365},
       number = {1854},
        pages = {1177-1178},
          doi = {10.1098/rsta.2006.1983},
       adsurl = {https://ui.adsabs.harvard.edu/abs/2007RSPTA.365.1177M}
}

@ARTICLE{2024_grb_deg,
       author = {{Garcia-Cifuentes}, Keneth and {Becerra}, Rosa Leticia and {De Colle}, Fabio and {Vargas}, Felipe},
        title = "{Unraveling parameter degeneracy in GRB data analysis}",
      journal = {\mnras},
         year = 2024,
        month = jan,
       volume = {527},
       number = {3},
        pages = {6752-6762},
          doi = {10.1093/mnras/stad3625},
archivePrefix = {arXiv},
       eprint = {2309.15825},
 primaryClass = {astro-ph.HE},
       adsurl = {https://ui.adsabs.harvard.edu/abs/2024MNRAS.527.6752G}
}

@ARTICLE{dessart_2016,
       author = {{Dessart}, Luc and {Hillier}, D. John and {Woosley}, Stan and {Livne}, Eli and {Waldman}, Roni and {Yoon}, Sung-Chul and {Langer}, Norbert},
        title = "{Inferring supernova IIb/Ib/Ic ejecta properties from light curves and spectra: correlations from radiative-transfer models}",
      journal = {\mnras},
         year = 2016,
        month = may,
       volume = {458},
       number = {2},
        pages = {1618-1635},
          doi = {10.1093/mnras/stw418},
archivePrefix = {arXiv},
       eprint = {1602.06280},
 primaryClass = {astro-ph.SR},
       adsurl = {https://ui.adsabs.harvard.edu/abs/2016MNRAS.458.1618D}
}

@ARTICLE{2004_Zeh,
       author = {{Zeh}, A. and {Klose}, S. and {Hartmann}, D.~H.},
        title = "{A Systematic Analysis of Supernova Light in Gamma-Ray Burst Afterglows}",
      journal = {\apj},
         year = 2004,
        month = jul,
       volume = {609},
       number = {2},
        pages = {952-961},
          doi = {10.1086/421100},
archivePrefix = {arXiv},
       eprint = {astro-ph/0311610},
 primaryClass = {astro-ph},
       adsurl = {https://ui.adsabs.harvard.edu/abs/2004ApJ...609..952Z}
}

@article{belkin_2024,
    author = {Belkin, S and Pozanenko, A S and Minaev, P Y and Pankov, N S and Volnova, A A and Rossi, A and Stratta, G and Benetti, S and Palazzi, E and Moskvitin, A S and Burhonov, O and Rumyantsev, V V and Klunko, E V and Inasaridze, R Ya and Reva, I V and Kim, V and Jelinek, M and Kann, D A and Volvach, A E and Volvach, L N and Xu, D and Zhu, Z and Fu, S and Mkrtchyan, A A},
    title = {GRB 201015A: from seconds to months of optical monitoring and supernova discovery},
    journal = {Monthly Notices of the Royal Astronomical Society},
    volume = {527},
    number = {4},
    pages = {11507-11520},
    year = {2024},
    month = {02},
    issn = {0035-8711},
    doi = {10.1093/mnras/stad3989},
    url = {https://doi.org/10.1093/mnras/stad3989},
    eprint = {https://academic.oup.com/mnras/article-pdf/527/4/11507/55118374/stad3989.pdf},
}

@article{Fulton_2023,
doi = {10.3847/2041-8213/acc101},
url = {https://doi.org/10.3847/2041-8213/acc101},
year = {2023},
month = {mar},
publisher = {The American Astronomical Society},
volume = {946},
number = {1},
pages = {L22},
author = {Fulton, M. D. and Smartt, S. J. and Rhodes, L. and Huber, M. E. and Villar, V. A. and Moore, T. and Srivastav, S. and Schultz, A. S. B. and Chambers, K. C. and Izzo, L. and Hjorth, J. and Chen, T.-W. and Nicholl, M. and Foley, R. J. and Rest, A. and Smith, K. W. and Young, D. R. and Sim, S. A. and Bright, J. and Zenati, Y. and de Boer, T. and Bulger, J. and Fairlamb, J. and Gao, H. and Lin, C.-C. and Lowe, T. and Magnier, E. A. and Smith, I. A. and Wainscoat, R. and Coulter, D. A. and Jones, D. O. and Kilpatrick, C. D. and McGill, P. and Ramirez-Ruiz, E. and Lee, K.-S. and Narayan, G. and Ramakrishnan, V. and Ridden-Harper, R. and Singh, A. and Wang, Q. and Kong, A. K. H. and Ngeow, C.-C. and Pan, Y.-C. and Yang, S. and Davis, K. W. and Piro, A. L. and Rojas-Bravo, C. and Sommer, J. and Yadavalli, S. K.},
title = {The Optical Light Curve of GRB 221009A: The Afterglow and the Emerging Supernova},
journal = {The Astrophysical Journal Letters}
}

@software{nessai,
       author = {{Williams}, Michael J. and {Veitch}, John and {Messenger}, Chris},
        title = "{nessai: Nested sampling with artificial intelligence}",
 howpublished = {Astrophysics Source Code Library, record ascl:2405.002},
         year = 2024,
        month = may,
          eid = {ascl:2405.002},
archivePrefix = {ascl},
       eprint = {2405.002},
       adsurl = {https://ui.adsabs.harvard.edu/abs/2024ascl.soft05002W}
}

@article{Jonker2013,
   title={DISCOVERY OF A NEW KIND OF EXPLOSIVE X-RAY TRANSIENT NEAR M86},
   volume={779},
   ISSN={1538-4357},
   url={http://dx.doi.org/10.1088/0004-637X/779/1/14},
   DOI={10.1088/0004-637x/779/1/14},
   journal={The Astrophysical Journal},
   publisher={American Astronomical Society},
   author={Jonker, P. G. and Glennie, A. and Heida, M. and Maccarone, T. and Hodgkin, S. and Nelemans, G. and Miller-Jones, J. C. A. and Torres, M. A. P. and Fender, R.},
   year={2013},
   month=nov, pages={14} }

@article{glennie2015,
    author = {Glennie, A. and Jonker, P. G. and Fender, R. P. and Nagayama, T. and Pretorius, M. L.},
    title = {Two fast X-ray transients in archival Chandra data},
    journal = {\mnras},
    volume = {450},
    number = {4},
    pages = {3765-3770},
    year = {2015},
    month = {05},
    issn = {0035-8711},
    doi = {10.1093/mnras/stv801},
    url = {https://doi.org/10.1093/mnras/stv801},
    eprint = {https://academic.oup.com/mnras/article-pdf/450/4/3765/5774833/stv801.pdf},
}

@ARTICLE{2024_Wichern,
       author = {{Wichern}, H.~C.~I. and {Ravasio}, M.~E. and {Jonker}, P.~G. and {Quirola-V{\'a}squez}, J.~A. and {Levan}, A.~J. and {Bauer}, F.~E. and {Kann}, D.~A.},
        title = "{Investigating the off-axis GRB afterglow scenario for extragalactic fast X-ray transients}",
      journal = {\aap},
         year = 2024,
        month = oct,
       volume = {690},
          eid = {A101},
        pages = {A101},
          doi = {10.1051/0004-6361/202450116},
archivePrefix = {arXiv},
       eprint = {2407.06371},
 primaryClass = {astro-ph.HE},
       adsurl = {https://ui.adsabs.harvard.edu/abs/2024A&A...690A.101W}
}

@article{2008D,
       author = {{Soderberg}, A.~M. and {Berger}, E. and {Page}, K.~L. and {Schady}, P. and {Parrent}, J. and {Pooley}, D. and {Wang}, X. -Y. and {Ofek}, E.~O. and {Cucchiara}, A. and {Rau}, A. and {Waxman}, E. and {Simon}, J.~D. and {Bock}, D.~C. -J. and {Milne}, P.~A. and {Page}, M.~J. and {Barentine}, J.~C. and {Barthelmy}, S.~D. and {Beardmore}, A.~P. and {Bietenholz}, M.~F. and {Brown}, P. and {Burrows}, A. and {Burrows}, D.~N. and {Byrngelson}, G. and {Cenko}, S.~B. and {Chandra}, P. and {Cummings}, J.~R. and {Fox}, D.~B. and {Gal-Yam}, A. and {Gehrels}, N. and {Immler}, S. and {Kasliwal}, M. and {Kong}, A.~K.~H. and {Krimm}, H.~A. and {Kulkarni}, S.~R. and {Maccarone}, T.~J. and {M{\'e}sz{\'a}ros}, P. and {Nakar}, E. and {O'Brien}, P.~T. and {Overzier}, R.~A. and {de Pasquale}, M. and {Racusin}, J. and {Rea}, N. and {York}, D.~G.},
        title = "{An extremely luminous X-ray outburst at the birth of a supernova}",
      journal = {\nat},
         year = 2008,
        month = may,
       volume = {453},
       number = {7194},
        pages = {469-474},
          doi = {10.1038/nature06997},
archivePrefix = {arXiv},
       eprint = {0802.1712},
 primaryClass = {astro-ph},
       adsurl = {https://ui.adsabs.harvard.edu/abs/2008Natur.453..469S}
}

@misc{martincarrillo2026,
      title={Failed jet breakout in the metal-poor broad-lined type Ic supernova 2026gzf}, 
      author={Antonio Martin-Carrillo and Christina C. Thöne and James K. Leung and Gregory Corcoran and Antonio de Ugarte Postigo and Peter G. Jonker and Luca Izzo and Andrew J. Levan and Benjamin P. Gompertz and Stéphane Basa and Nikhil Sarin and Jonathan Quirola-Vásquez and Rob A. J. Eyles-Ferris and Riccardo Brivio and Alan M. Watson and Laura Cotter and Jennifer Alexandra Chacón and Andrea Rossi and Andrea Melandri and Piramon Kumnurdmanee and Nial R. Tanvir and Anshika Gupta and Franz E. Bauer and Jean-Grégoire Ducoin and Andrea Reguitti and Kuntal Misra and Dong Xu and Susanna D. Vergani and Wen-fai Fong and Kendall Ackley and Edilberto Aguilar-Ruiz and Dalya Akl and Miguel Ángel Aloy and Jie An and Camila Angulo-Valdez and Sarah Antier and Jean-Luc Atteia and Rosa L. Becerra and Rene P. Breton and Nathaniel R. Butler and Sergio Campana and Francesco Carotenuto and Jorge Casares Velázquez and Ashley A. Chrimes and Valerio D'Elia and Joyce N. D. van Dalen and Fabio De Colle and Massimiliano De Pasquale and Vik S. Dhillon and Damien Dornic and Martin J. Dyer and Matteo Ferro and Morgan Fraser and Andrew S. Fruchter and Francis Fortin and Duncan K. Galloway and Leonardo García-García and Stefan Geier and Ramandeep Gill and Noémie Globus and Roberto Gualandi and Marion Guelfand and Francesco Guidolin and Dieter H. Hartmann and Agnes P. C. van Hoof and Pall Jakobsson and Divyanshu Janghel and Tom L. Killestein and Sylvio Klose and Shiho Kobayashi and Rubina Kotak and Amit Kumar and Asuka Kuwata and Tanmoy Laskar and William H. Lee and Massimiliano Lincetto and Gianluca Lombardi and Diego López-Cámara and Joseph D. Lyman and Elisabetta Maiorano and Keiichi Maeda and Nikos Mandarakas and Francesco Magnani and Jirong Mao and Enrique Moreno Méndez and Ana María Nicuesa Guelbenzu and Kanthanakorn Noysena and Laura K. Nuttall and Paul T. O'Brien and David O'Neill and Paolo Ochner and Margarita Pereyra and Giovanna Pugliese and Gavin Ramsay and Lauren Rhodes and Andrea Saccardi and Ruben Salvaterra and Fredd Sánchez Álvarez and Benjamin Schneider and Steve Schulze and Rhaana L. C. Starling and Danny Steeghs and Kzrysztof Ulaczyk and Chiara Ventura and Tayyaba Zafar and Zi-Pei Zhu},
      year={2026},
      eprint={2606.10002},
      archivePrefix={arXiv},
      primaryClass={astro-ph.HE},
      url={https://arxiv.org/abs/2606.10002}, 
}

@misc{oconnor2026,
      title={EP260321a/SN 2026gzf: The Faintest Shock Breakout Associated with a Broad-Lined Supernova}, 
      author={Brendan O'Connor and Xander J. Hall and Malte Busmann and Daniel Gruen and Alberto Floris and Tomas Cabrera and Ziyuan Zhu and Antonella Palmese and Dylan Green and John Banovetz and Julius Gassert and Christopher L. Fryer and Roberto Ricci and Eleonora Troja and Surya Shivaprasad and Gregory R. Zeimann and Ariel J. Amsellem and Stephen Bailey and Segev BenZvi and Simone Dichiara and Hendrik van Eerten and Jeremy Hare and Lei Hu and Christopher M. Irwin and Keerthi Kunnumkai and Konstantin Malanchev and Mitra Maleki and Michael J. Moss and Adam D. Myers and Dheeraj Pasham and Christoph Ries and Geoffrey Ryan and David Schlegel and Michael Schmidt and Silona Wilke and Yu-Han Yang},
      year={2026},
      eprint={2606.09992},
      archivePrefix={arXiv},
      primaryClass={astro-ph.HE},
      url={https://arxiv.org/abs/2606.09992}, 
}

@misc{yuan2026,
      title={X-rays breaking out of pre-explosion ejecta mark a supernova's first light}, 
      author={Weimin Yuan and Qiu-Ju Huang and Jin-Ping Zhu and Yun-Wei Yu and Dong Xu and Chen Zhang and Zhuo Li and Yuan Liu and Tao An and Giulia Gianfagna and Weikang Zheng and Guowang Du and Xing Liu and Ji-An Jiang and Johan P. U. Fynbo and Alexei S. Pozanenko and Junjie Jin and Yi Yang and Jinsong Deng and Hui Sun and Guang-Lei Wu and Yu-Hao Zhang and Bao Wang and Yu Wang and Xiangyu Wang and Bin-Bin Zhang and Yong Chen and Yonghe Zhang and Bo Wang and Xiaofeng Wang and Xuefeng Wu and Zigao Dai and Jie An and G. C. Anupama and Arvind Balasubramanian and Congying Bao and Aru Beri and Varun Bhalerao and Thomas G. Brink and Gabriele Bruni and Minxuan Cai and Zhiming Cai and Krittapas Chanchaiworawit and Yehai Chen and Huaqing Cheng and Bertrand Cordier and Chenzhou Cui and Weiwei Cui and Cuiyuan Dai and D. Eappachen and M. V. Eselevich and Xiao Fan and Zhou Fan and Yuan Fang and Hua Feng and Alexei V. Filippenko and Shaoyu Fu and He Gao and Jinjun Geng and Vitaly Goranskij and Ju Guan and Dawei Han and Jinxin Hao and Linbo He and Min He and Jingwei Hu and Maohai Huang and Shumei Jia and Ziqing Jia and Shuaiqing Jiang and Chichuan Jin and Ge Jin and Peter Jonker and E. V. Klunko and Albert K. H. Kong and Chengkui Li and Dongyue Li and Rui-Zhi Li and Wenxiong Li and Run-Duo Liang and Zhixing Ling and Congzhan Liu and Huaqiu Liu and Liangduan Liu and Xiangkun Liu and Xiaowei Liu and Yuanqi Liu and Zhengwei Liu and Fangjun Lu and Jirong Mao and Xuan Mao and A. S. Moskvitin and Haiyang Mu and Kirpal Nandra and Jan-Uwe Ness and Kangrui Ni and Kanthanakorn Noysena and Paul O'Brien and Haiwu Pan and Yu Pan and N. S. Pankov and Luigi Piro and J. Quirola-Vasquez and Arne Rau and Nanda Rea and D. K. Sahu and Aditya Pawan Saikia and Jeremy Sanders and Liming Song and Olga Spiridonova and Ning-Chen Sun and Shengli Sun and Xiaojin Sun and Yuyin Tan and Aishwarya Linesh Thakur and Samaporn Tinyanont and Valery Vlasyuk and A. V. Volnova and Ailing Wang and Hong Wu and Qianrui Wu and Haitao Xu and Zelin Xu and Changbin Xue and Yi-Han Iris Yin and I. A. Zaznobin and Jia-Sen Zhang and Shuang-Nan Zhang and Songbo Zhang and Yu Zhang and Zipei Zhu and Zecheng Zou and Bing Zhang},
      year={2026},
      eprint={2606.10014},
      archivePrefix={arXiv},
      primaryClass={astro-ph.HE},
      url={https://arxiv.org/abs/2606.10014}, 
}

@ARTICLE{rastinejad2026,
       author = {{Rastinejad}, Jillian C. and {Srinivasaragavan}, Gokul and {Sarin}, Nikhil and {O'Dwyer}, Tanner and {Cenko}, S. Bradley and {Leung}, James K. and {Nugent}, Anya E. and {Perley}, Daniel A. and {Schroeder}, Genevieve and {Anand}, Shreya and {Ahumada}, Tomas and {Andreoni}, Igor and {Bochenek}, Aleksandra and {Corsi}, Alessandra and {Fremling}, Christoffer and {Ho}, Anna Y.~Q. and {Kasliwal}, Mansi M. and {Mo}, Geoffrey and {Salgundi}, Anirudh and {Sippy}, Kendall I. and {Sollerman}, J. and {Bellm}, Eric C. and {Chen}, Tracy X. and {Coughlin}, Michael W. and {Davis}, Michael C. and {De Colle}, Fabio and {Frostig}, Danielle and {Fryer}, Christopher L. and {Graham}, Michael J. and {Hall}, Xander J. and {Hinds}, K.-R. and {Izzo}, Luca and {Jacobson-Galan}, Wynn and {Lourie}, Nathan P. and {Maeda}, Keiichi and {Purdum}, Josiah and {Rusholme}, Ben and {Singh}, Avinash and {Stein}, Robert},
        title = "{A Multi-Wavelength View of the First Type Ic-BL Supernova with an Einstein Probe X-ray Shock Breakout}",
      journal = {arXiv e-prints},
         year = 2026,
        month = jun,
          eid = {arXiv:2606.10011},
        pages = {arXiv:2606.10011},
          doi = {10.48550/arXiv.2606.10011},
archivePrefix = {arXiv},
       eprint = {2606.10011},
 primaryClass = {astro-ph.HE},
       adsurl = {https://ui.adsabs.harvard.edu/abs/2026arXiv260610011R}
}

@ARTICLE{chen2026,
       author = {{Chen}, Ting-Wan and {Aryan}, Amar and {Yang}, Sheng and {Smartt}, Stephen J. and {Moriya}, Takashi J. and {Brennan}, Se{\'a}n J. and {Stritzinger}, Maximilian D. and {Martin}, Bailey and {Nicholl}, Matt and {Kong}, Albert K.~H. and {Gillanders}, James H. and {Dutta}, Anirban and {Schmidt}, Brian P. and {Cheng}, Yu-Chi and {Huber}, Mark E. and {Lai}, Cheng-Han and {Lee}, Chien-Hsiu and {Lee}, Yu-Hsing and {Ngeow}, Chow-Choong and {Smith}, Ken W. and {Ashall}, Christopher and {Auchettl}, Katie and {Burns}, Chris R. and {Chambers}, Kenneth C. and {Chen}, Zhi-Yue and {de Boer}, Thomas and {Hsiao}, Eric Y. and {Ngo Thanh Ho}, Khoa and {Hoogendam}, Willem B. and {Jones}, David O. and {Kankare}, Erkki and {Killestein}, Tom L. and {Kuncarayakti}, Hanindyo and {Lee}, Meng-Han and {Li}, Chuan-Jui and {Lin}, Chien-Cheng and {Lidman}, Christopher and {Lowe}, Thomas B. and {Magnier}, Eugene A. and {Medler}, Kyle and {M{\"o}ller}, Anais and {Moore}, Thomas and {Morrell}, Nidia and {Paek}, Gregory S.~H. and {Pfeffer}, Cameron M. and {Qiang}, Da-Chun and {Rauf}, Liana and {Reynolds}, Thomas M. and {Sankar. K}, Aiswarya and {Srivastav}, Shubham and {Tweddle}, Jack and {Wainscoat}, Richard and {Wang}, Ze-Ning and {Xiao}, Huangfei and {Zhu}, Zonghong},
        title = "{Decadal pre-explosion activity and circumstellar interaction in a supernova}",
      journal = {arXiv e-prints},
         year = 2026,
        month = jun,
          eid = {arXiv:2606.10009},
        pages = {arXiv:2606.10009},
          doi = {10.48550/arXiv.2606.10009},
archivePrefix = {arXiv},
       eprint = {2606.10009},
 primaryClass = {astro-ph.HE},
       adsurl = {https://ui.adsabs.harvard.edu/abs/2026arXiv260610009C}
}

@article{cotter2026n,
      title={Probing a new subclass of llGRB-SN transients: Insights from EP250304a and its associated supernova}, 
      author={L. Cotter and A. Martin-Carrillo and R. A. J. Eyles-Ferris and L. Izzo and D. B. Malesani and Y. Julakanti and G. Corcoran and A. Saccardi and P. G. Jonker and A. J. Levan and F. Carotenuto and P. T. O'Brien and J. H. Gillanders and J. N. D. van Dalen and M. E. Ravasio and S. Schulze and N. Sarin and F. E. Bauer and M. Fraser and J. Quirola-Vasquez and A. P. C. van Hoof and S. J. Smartt and C. Gall and A. Rest and C. T. Murphey and N. Tanvir and T. -W. Chen and S. Campana and C. Ashall and J. P. Anderson and J. A. Chacon and F. J. Cowie and V. D'Elia and L. Galbany and C. P. Gutierrez and D. H. Hartmann and P. Jakobsson and S. Kobayashi and A. H. Kong and P. Mazalli and T. E. Muller-Bravo and M. De Pasquale and L. Rhodes and A. Rossi and J. Sanchez-Sierras and J. Sollerman and A. Andersson and A. Aryan and T. de Boer and J. S. Bright and K. C. Chambers and M. Gromadzki and M. E. Huber and C. Inserra and T. Lowe and P. Minguez and G. S. Narayan and M. Nicholl and G. S. H. Paek and A. Sedgewick and K. W. Smith and J. W. Tweddle and S. Yang},
      year={2026},
      journal = {ArXiv e-prints},
      eprint={2606.06213},
      archivePrefix={arXiv},

      primaryClass={astro-ph.HE},
      url={https://arxiv.org/abs/2606.06213}, 
}

@misc{0827b,
      title={EP250827b/SN 2025wkm: An X-ray Flash-Supernova Powered by a Central Engine and Circumstellar Interaction}, 
      author={Gokul P. Srinivasaragavan and Dongyue Li and Xander J. Hall and Ore Gottlieb and Genevieve Schroeder and Heyang Liu and Brendan O'Connor and Chichuan Jin and Mansi Kasliwal and Tomás Ahumada and Qinyu Wu and Christopher L. Fryer and Annabelle E. Niblett and Dong Xu and Maria Edvige Ravasio and Grace Daja and Wenxiong Li and Shreya Anand and Anna Y. Q. Ho and Hui Sun and Daniel A. Perley and Lin Yan and Eric Burns and S. Bradley Cenko and Jesper Sollerman and Nikhil Sarin and Anthony L. Piro and Amar Aryan and M. Coleman Miller and Jie An and Tao An and Moira Andrews and Jule Augustin and Eric C. Bellm and Aleksandra Bochenek and Malte Busmann and Krittapas Chanchaiworawit and Huaqing Chen and Maria D. Caballero-García and Alberto J. Castro-Tirado and Ali Esamdin and Jennifer Faba-Moreno and Joseph Farah and Emilio Fernández-García and Shaoyu Fu and Johan P. U. Fynbo and Julius Gassert and Estefania Padilla Gonzalez and Ignacio Pérez-García and Matthew Graham and Maria Gritsevich and Daniel Gruen and Sergiy Guziy and D. Andrew Howell and Linbo He and Jingwei Hu and You-Dong Hu and Abdusamatjan Iskandar and Joahan Castaneda Jaims and Ji-An Jiang and Ning Jiang and Shuaijiao Jiang and Runduo Liang and Zhixing Ling and Jialian Liu and Xing Liu and Yuan Liu and Frank J. Masci and Curtis McCully and Megan Newsome and Kanthanakorn Noysena and Shashi B. Pandey and Kangrui Ni and Antonella Palmese and Han-Long Peng and Josiah Purdum and Yu-Jing Qin and Sam Rose and Ben Rusholme and Rubén Sánchez-Ramírez and Cassie Sevilla and Roger Smith and Yujia Song and Niharika Sravan and Robert Stein and Constantin Tabor and Giacomo Terreran and Samaporn Tinyanont and Pablo Vega and Letian Wang and Tinggu Wang and Xiaofeng Wang and Siyu Wu and Xuefeng Wu and Kathryn Wynn and Yunfei Xu and Shengyu Yan and Weimin Yuan and Binbin Zhang and Chen Zhang and Zipei Zhu and Xiaoxiong Zuo and Gursimran Bhullar},
      year={2026},
      eprint={2512.10239},
      archivePrefix={arXiv},
      primaryClass={astro-ph.HE},
      url={https://arxiv.org/abs/2512.10239}, 
}

@ARTICLE{Garcia_2024,
       author = {{Garcia-Cifuentes}, Keneth and {Becerra}, Rosa Leticia and {De Colle}, Fabio and {Vargas}, Felipe},
        title = "{Unraveling parameter degeneracy in GRB data analysis}",
      journal = {\mnras},
         year = 2024,
        month = jan,
       volume = {527},
       number = {3},
        pages = {6752-6762},
          doi = {10.1093/mnras/stad3625},
archivePrefix = {arXiv},
       eprint = {2309.15825},
 primaryClass = {astro-ph.HE},
       adsurl = {https://ui.adsabs.harvard.edu/abs/2024MNRAS.527.6752G}
}

@article{1999_Sari,
   title={Jets in Gamma-Ray Bursts},
   volume={519},
   ISSN={0004-637X},
   url={http://dx.doi.org/10.1086/312109},
   DOI={10.1086/312109},
   number={1},
   journal={The Astrophysical Journal},
   publisher={American Astronomical Society},
   author={Sari, Re’em and Piran, Tsvi and Halpern, J. P.},
   year={1999},
   month=July, pages={L17–L20} }

@article{Wu_2004,
   title={Jet Break Time–Flux Density Relationship and Constraints on Physical Parameters of Gamma‐Ray Burst Afterglows},
   volume={615},
   ISSN={1538-4357},
   url={http://dx.doi.org/10.1086/424378},
   DOI={10.1086/424378},
   number={1},
   journal={The Astrophysical Journal},
   publisher={American Astronomical Society},
   author={Wu, X. F. and Dai, Z. G. and Liang, E. W.},
   year={2004},
   month=Nov, pages={359–365} }

@article{Zhang_magnetar,
       author = {{Zhang}, Bing},
        title = "{Early X-Ray and Optical Afterglow of Gravitational Wave Bursts from Mergers of Binary Neutron Stars}",
      journal = {\apjl},
         year = 2013,
        month = jan,
       volume = {763},
       number = {1},
          eid = {L22},
        pages = {L22},
          doi = {10.1088/2041-8205/763/1/L22},
archivePrefix = {arXiv},
       eprint = {1212.0773},
 primaryClass = {astro-ph.HE},
       adsurl = {https://ui.adsabs.harvard.edu/abs/2013ApJ...763L..22Z}
}

@article{2024Vasquez,
       author = {{Quirola-V{\'a}squez}, J. and {Bauer}, F.~E. and {Jonker}, P.~G. and {Brandt}, W.~N. and {Eappachen}, D. and {Levan}, A.~J. and {L{\'o}pez}, E. and {Luo}, B. and {Ravasio}, M.~E. and {Sun}, H. and {Xue}, Y.~Q. and {Yang}, G. and {Zheng}, X.~C.},
        title = "{Probing a magnetar origin for the population of extragalactic fast X-ray transients detected by Chandra}",
      journal = {\aap},
         year = 2024,
        month = mar,
       volume = {683},
          eid = {A243},
        pages = {A243},
          doi = {10.1051/0004-6361/202347629},
archivePrefix = {arXiv},
       eprint = {2401.01415},
 primaryClass = {astro-ph.HE},
       adsurl = {https://ui.adsabs.harvard.edu/abs/2024A&A...683A.243Q}
}

@article{2026_zheng,
       author = {{Zheng}, Jian-He and {Lu}, Wenbin},
        title = "{Fast X-Ray Transients Produced by Off-axis Jet Cocoons from Long Gamma-Ray Bursts}",
      journal = {\apjl},
         year = 2026,
        month = may,
       volume = {1003},
       number = {1},
          eid = {L19},
        pages = {L19},
          doi = {10.3847/2041-8213/ae67f2},
archivePrefix = {arXiv},
       eprint = {2603.09674},
 primaryClass = {astro-ph.HE},
       adsurl = {https://ui.adsabs.harvard.edu/abs/2026ApJ..1003L..19Z}
}

@misc{sarin2026,
      title={Signatures of $^{56}$Ni Mixing and Neutron-rich Ejecta in Supernovae}, 
      author={Nikhil Sarin},
      year={2026},
      eprint={2606.23780},
      archivePrefix={arXiv},
      primaryClass={astro-ph.HE},
      url={https://arxiv.org/abs/2606.23780}, 
}

@ARTICLE{barnes_2018, 
    title= "{A GRB and Broad-lined Type Ic Supernova from a Single Central Engine}", 
    volume={860}, 
    DOI={10.3847/1538-4357/aabf84}, 
    number={1}, 
    journal={The Astrophysical Journal}, 
    author={Barnes, Jennifer and Duffell, Paul C. and Liu, Yuqian and Modjaz, Maryam and Bianco, Federica B. and Kasen, Daniel and MacFadyen, Andrew I.}, 
    year={2018}, 
    pages={38}}

@article{dessart_2012,
       author = {{Dessart}, Luc and {Hillier}, D. John and {Li}, Chengdong and {Woosley}, Stan},
        title = "{On the nature of supernovae Ib and Ic}",
      journal = {\mnras},
         year = 2012,
        month = aug,
       volume = {424},
       number = {3},
        pages = {2139-2159},
          doi = {10.1111/j.1365-2966.2012.21374.x},
archivePrefix = {arXiv},
       eprint = {1205.5349},
 primaryClass = {astro-ph.SR},
       adsurl = {https://ui.adsabs.harvard.edu/abs/2012MNRAS.424.2139D}
}

@ARTICLE{Woosley_2002,
       author = {{Woosley}, S.~E. and {Heger}, A. and {Weaver}, T.~A.},
        title = "{The evolution and explosion of massive stars}",
      journal = {Reviews of Modern Physics},
         year = 2002,
        month = nov,
       volume = {74},
       number = {4},
        pages = {1015-1071},
          doi = {10.1103/RevModPhys.74.1015},
       adsurl = {https://ui.adsabs.harvard.edu/abs/2002RvMP...74.1015W}
}

\appendix
\section{Additional Figures}
\label{app:tables}

\begin{figure}[h]
\centering
\includegraphics[width=0.6\linewidth]{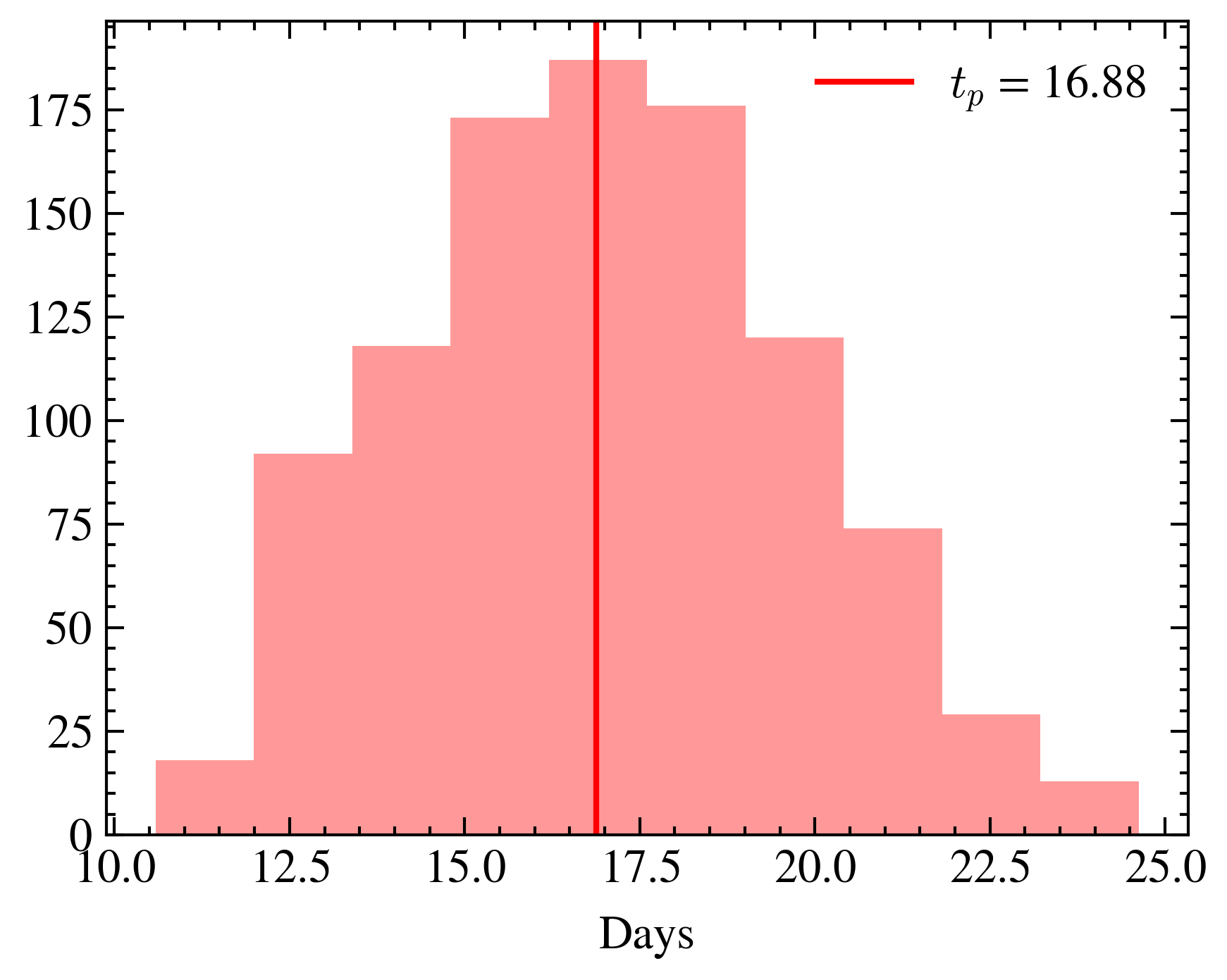}
\caption{Distribution of SN peak times in the $r$-band data calculated from the 1000 simulated parameters.}
\label{rband_peak}
\end{figure}

\begin{figure}[h]
\centering
\includegraphics[width=0.6\linewidth]{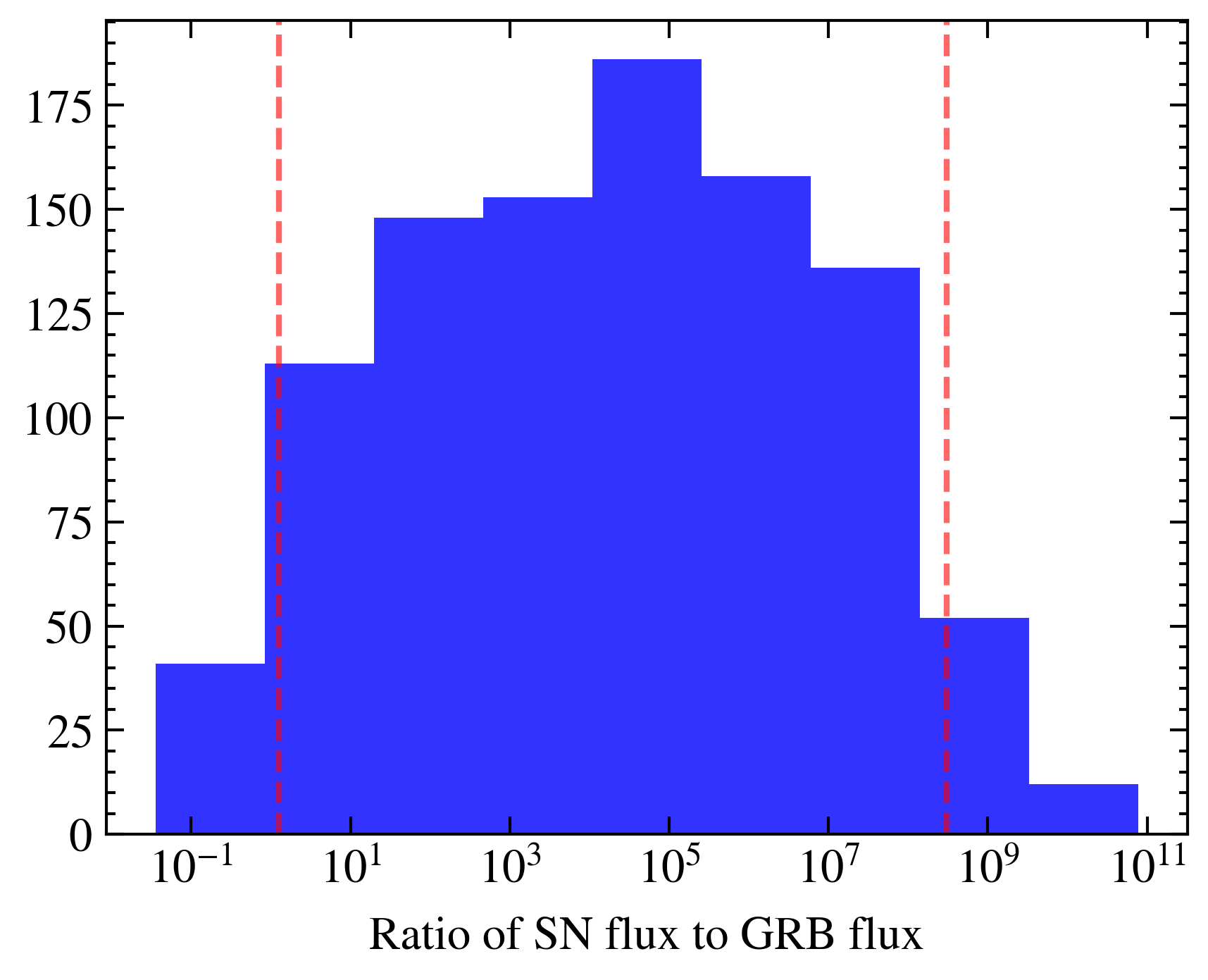}
\caption{Distribution of the ratio between the SN and GRB flux in the $r$-band at the median SN peak time of 16.88 days. The red dashed lines represent the lower and upper ends of the 90\% credible interval, respectively.}
\label{grbsn_ratio}
\end{figure}

\begin{figure}[h]
\centering
\includegraphics[width=0.6\linewidth]{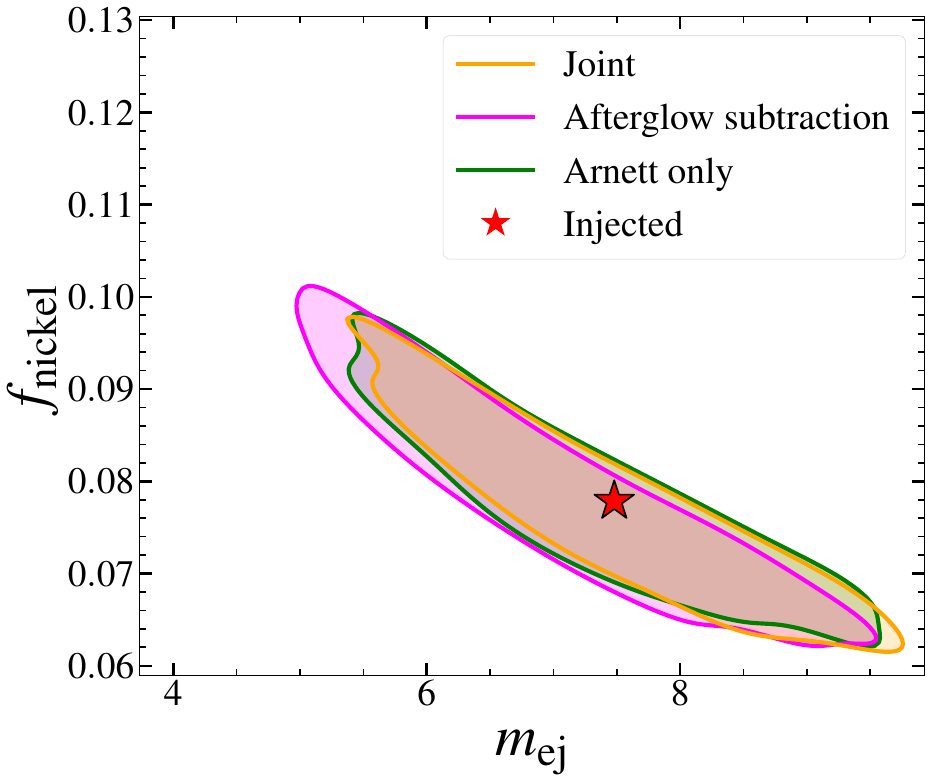}
\caption{Relationship between the $m_\text{ej}$ and $f_\text{nickel}$ parameters for the SN dominated case with an early jet break using the posteriors obtained from each modelling approach. Each region represents the $1\sigma$ credible interval of the posterior.}
\label{early_sndom_fnimej}
\end{figure}

\begin{figure}[h]
\centering
\includegraphics[width=0.6\linewidth]{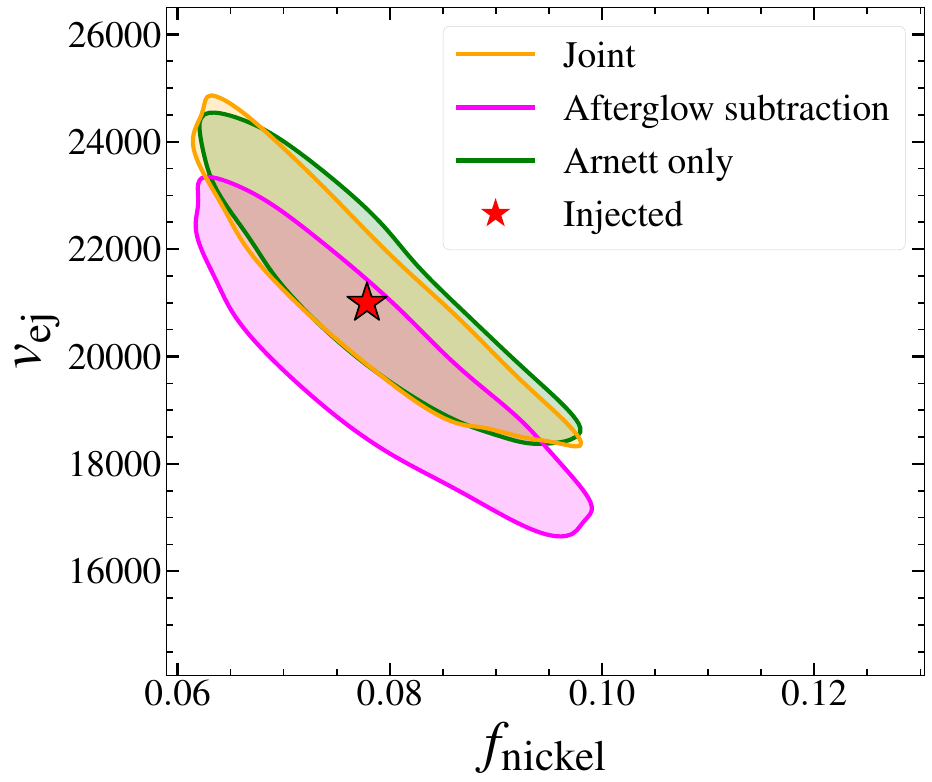}
\caption{Relationship between the $f_\text{nickel}$ and $v_\text{ej}$ parameters for the SN dominated case with an early jet break using the posteriors obtained from each modelling approach. Each region represents the $1\sigma$ credible interval of the posterior.}
\label{early_sndom_fnimej}
\end{figure}

\begin{figure}[h]
\centering
\includegraphics[width=0.6\linewidth]{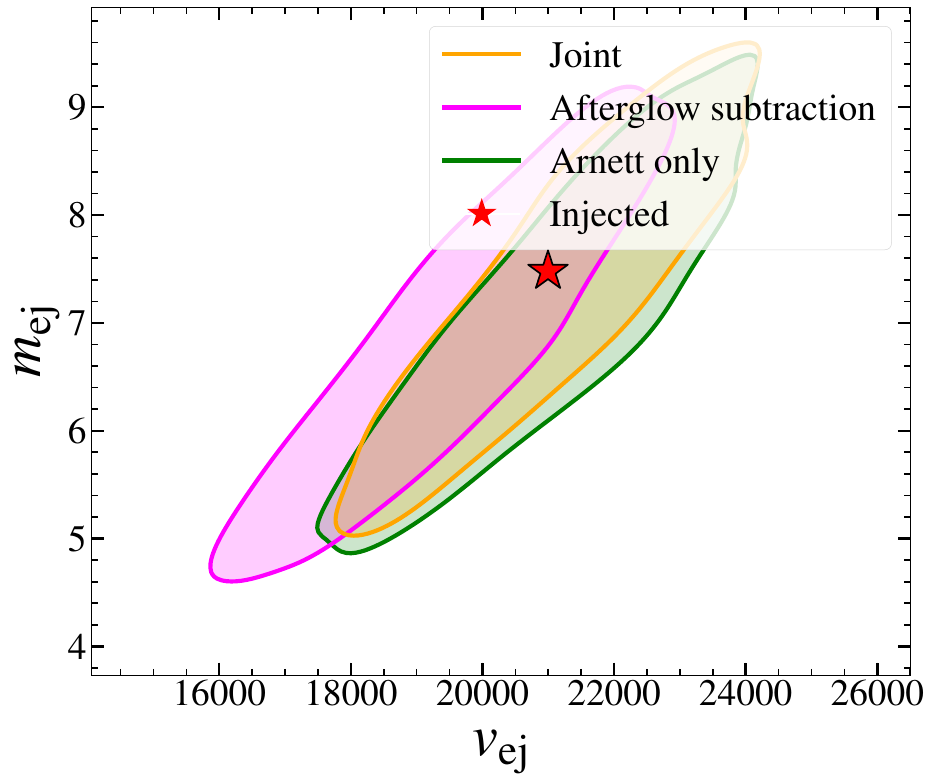}
\caption{Relationship between the $v_\text{ej}$ and $m_\text{ej}$ parameters for the SN dominated case with an early jet break using the posteriors obtained from each modelling approach. Each region represents the $1\sigma$ credible interval of the posterior.}
\label{early_sndom_vejmej}
\end{figure}

\begin{figure}[h]
\centering
\includegraphics[width=0.6\linewidth]{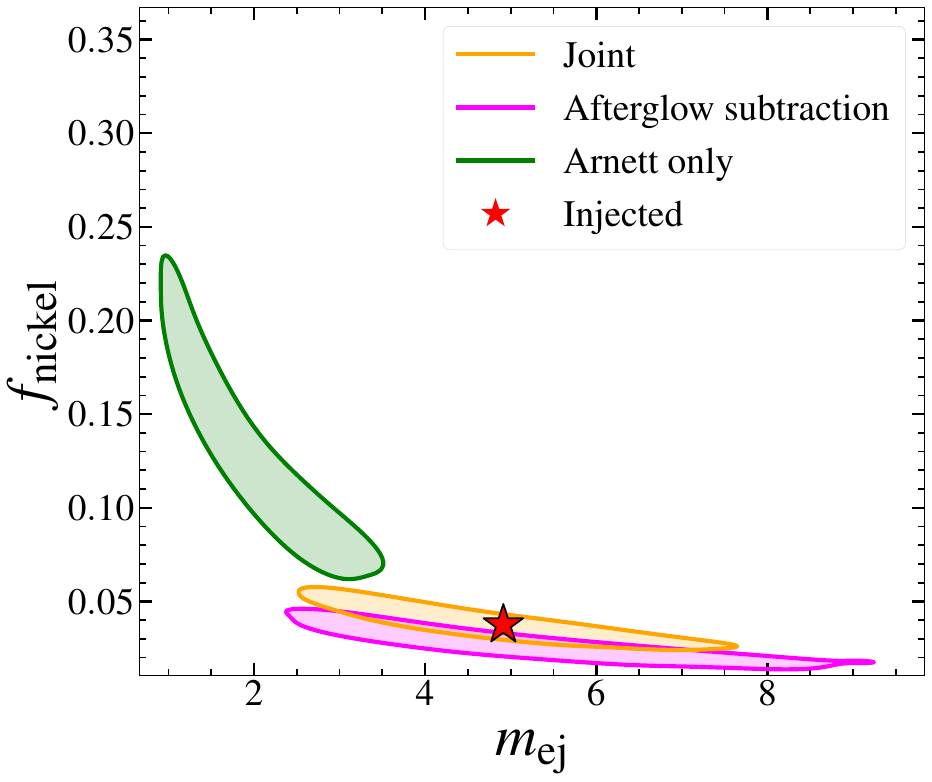}
\caption{Relationship between the $m_\text{ej}$ and $f_\text{nickel}$ parameters for the GRB dominated case with an early jet break using the posteriors obtained from each modelling approach. Each region represents the $1\sigma$ credible interval of the posterior.}
\label{early_grbdom_fnimej}
\end{figure}

\begin{figure}[h]
\centering
\includegraphics[width=0.6\linewidth]{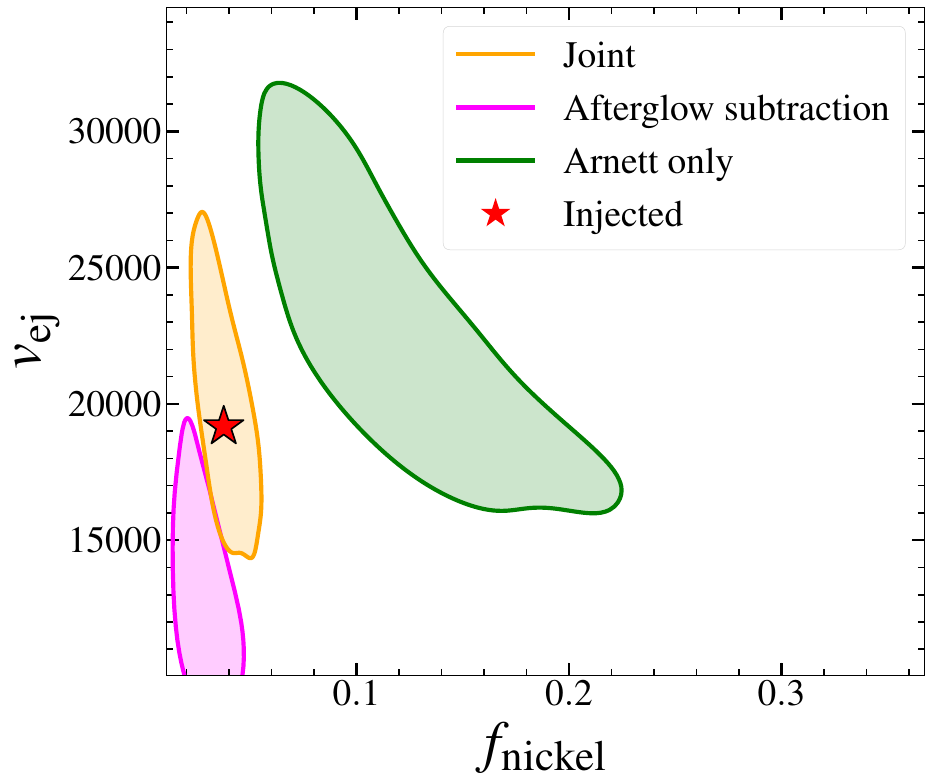}
\caption{Relationship between the $f_\text{nickel}$ and $v_\text{ej}$ parameters for the GRB dominated case with an early jet break using the posteriors obtained from each modelling approach. Each region represents the $1\sigma$ credible interval of the posterior.}
\label{early_grbdom_fnimej}
\end{figure}

\begin{figure}[h]
\centering
\includegraphics[width=0.6\linewidth]{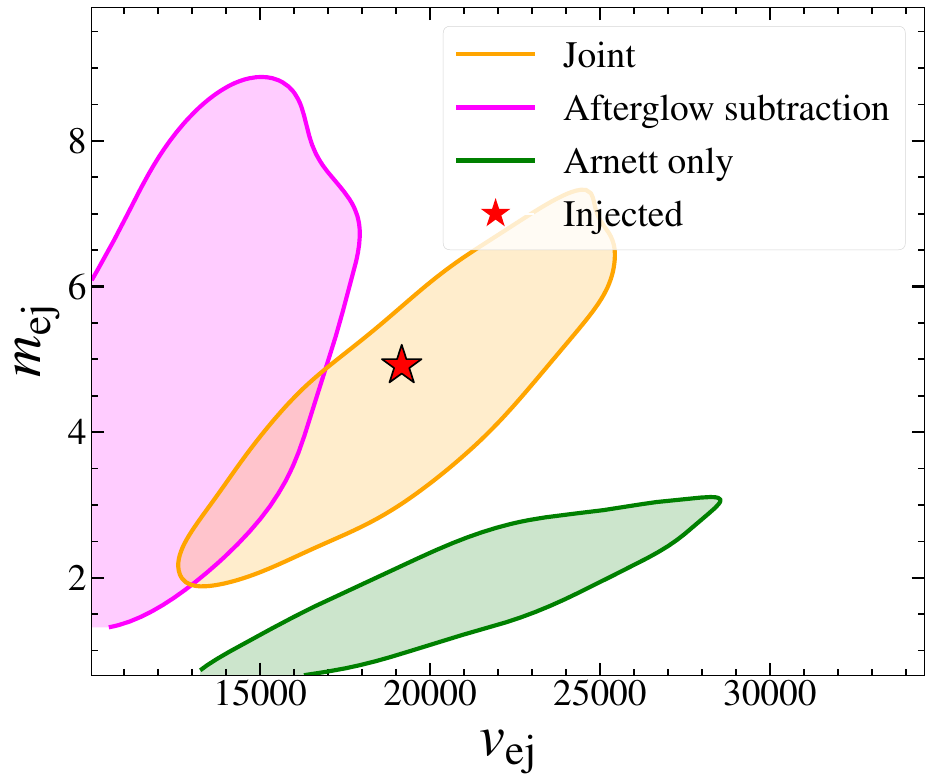}
\caption{Relationship between the $v_\text{ej}$ and $m_\text{ej}$ parameters for the GRB dominated case with an early jet break using the posteriors obtained from each modelling approach. Each region represents the $1\sigma$ credible interval of the posterior.}
\label{early_grbdom_vejmej}
\end{figure}

\begin{figure}[h]
\centering
\includegraphics[width=0.6\linewidth]{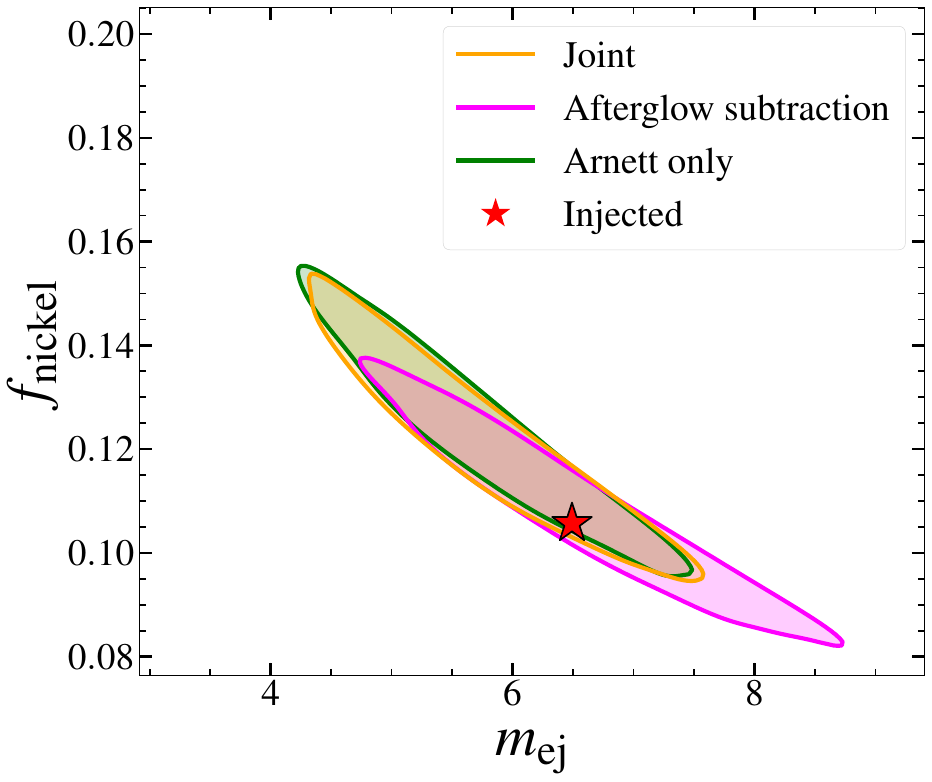}
\caption{Relationship between the $m_\text{ej}$ and $f_\text{nickel}$ parameters for the  normal GRB-SN case with an early jet break using the posteriors obtained from each modelling approach. Each region represents the $1\sigma$ credible interval of the posterior.}
\label{early_normal_fnimej}
\end{figure}

\begin{figure}[h]
\centering
\includegraphics[width=0.6\linewidth]{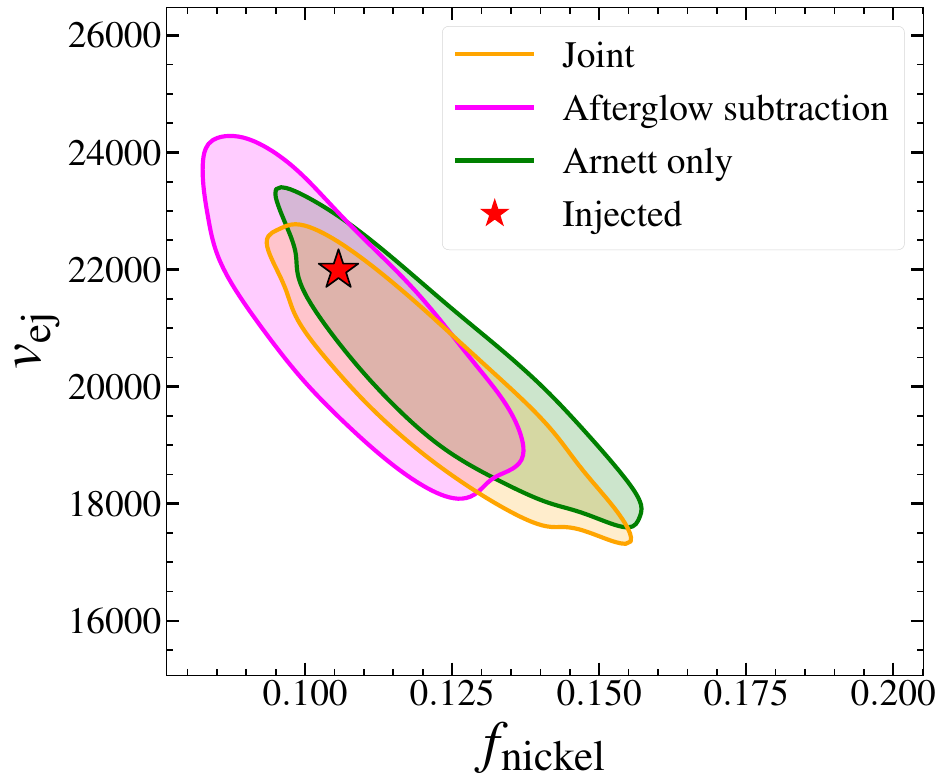}
\caption{Relationship between the $f_\text{nickel}$ and $v_\text{ej}$ parameters for the normal GRB-SN case with an early jet break using the posteriors obtained from each modelling approach. Each region represents the $1\sigma$ credible interval of the posterior.}
\label{early_normal_fnimej}
\end{figure}

\begin{figure}[h]
\centering
\includegraphics[width=0.6\linewidth]{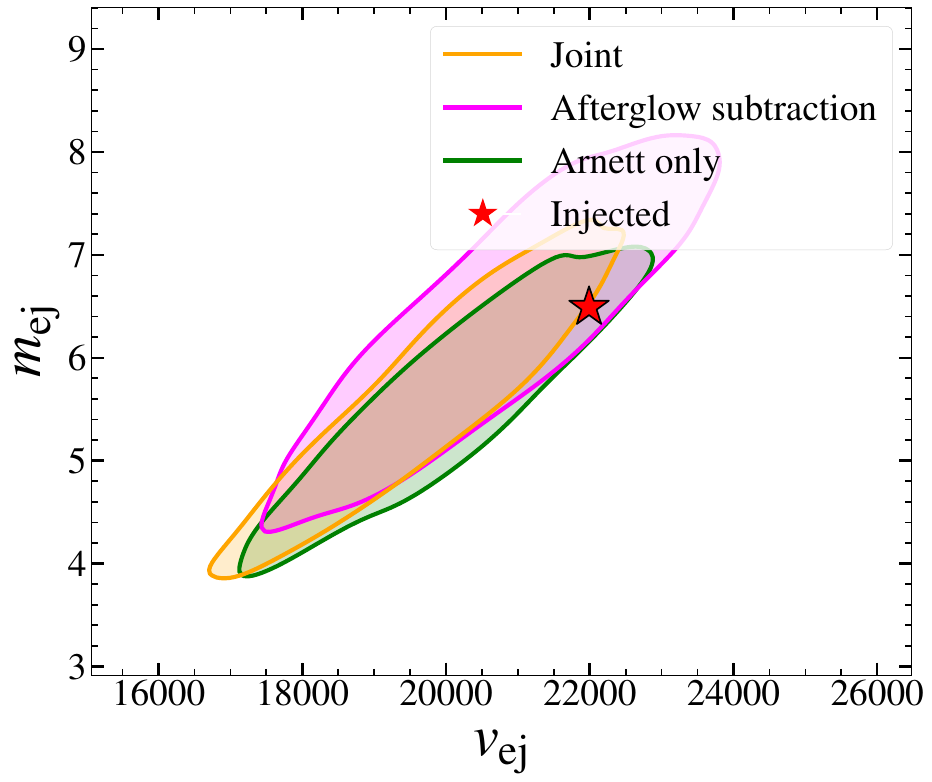}
\caption{Relationship between the $v_\text{ej}$ and $m_\text{ej}$ parameters for the normal GRB-SN case with an early jet break using the posteriors obtained from each modelling approach. Each region represents the $1\sigma$ credible interval of the posterior.}
\label{early_normal_vejmej}
\end{figure}

\begin{figure}[h]
\centering
\includegraphics[width=0.6\linewidth]{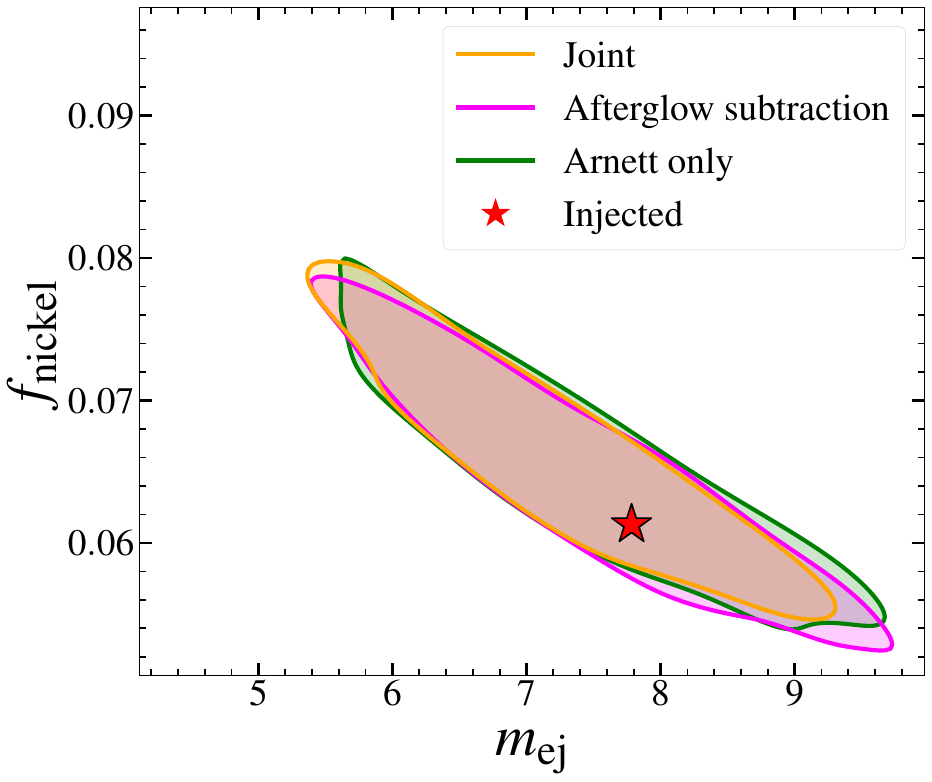}
\caption{Relationship between the $m_\text{ej}$ and $f_\text{nickel}$ parameters for the SN dominated case with an late jet break using the posteriors obtained from each modelling approach. Each region represents the $1\sigma$ credible interval of the posterior.}
\label{late_sndom_fnimej}
\end{figure}

\begin{figure}[h]
\centering
\includegraphics[width=0.6\linewidth]{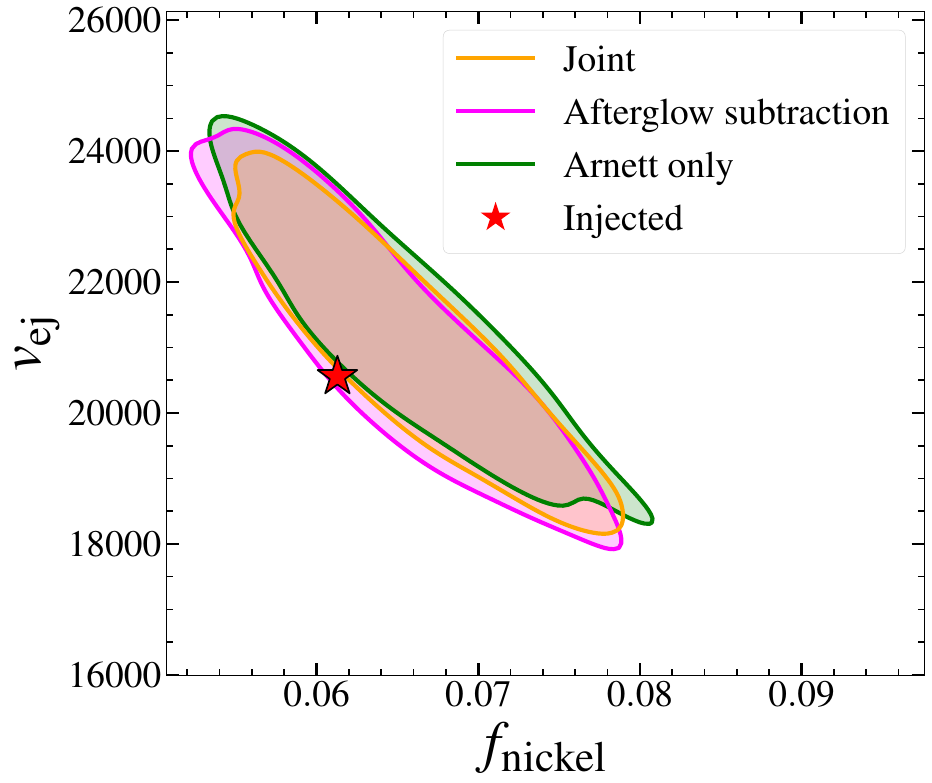}
\caption{Relationship between the $f_\text{nickel}$ and $v_\text{ej}$ parameters for the SN dominated case with an late jet break using the posteriors obtained from each modelling approach. Each region represents the $1\sigma$ credible interval of the posterior.}
\label{late_sndom_fnimej}
\end{figure}

\begin{figure}[h]
\centering
\includegraphics[width=0.6\linewidth]{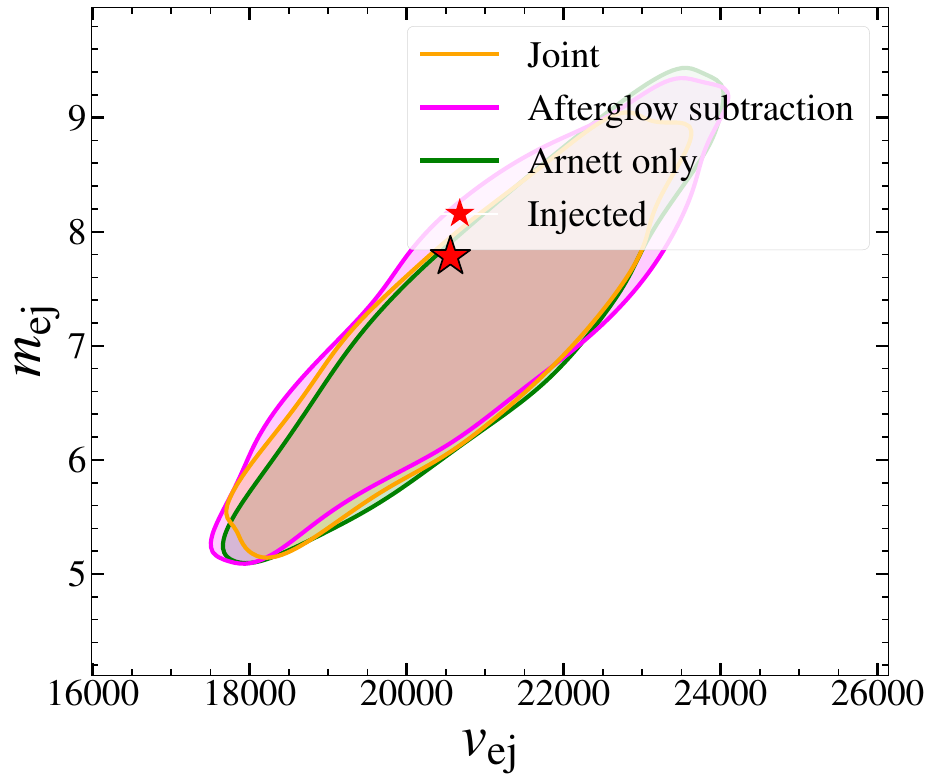}
\caption{Relationship between the $v_\text{ej}$ and $m_\text{ej}$ parameters for the SN dominated case with an late jet break using the posteriors obtained from each modelling approach. Each region represents the $1\sigma$ credible interval of the posterior.}
\label{late_sndom_vejmej}
\end{figure}

\begin{figure}[h]
\centering
\includegraphics[width=0.6\linewidth]{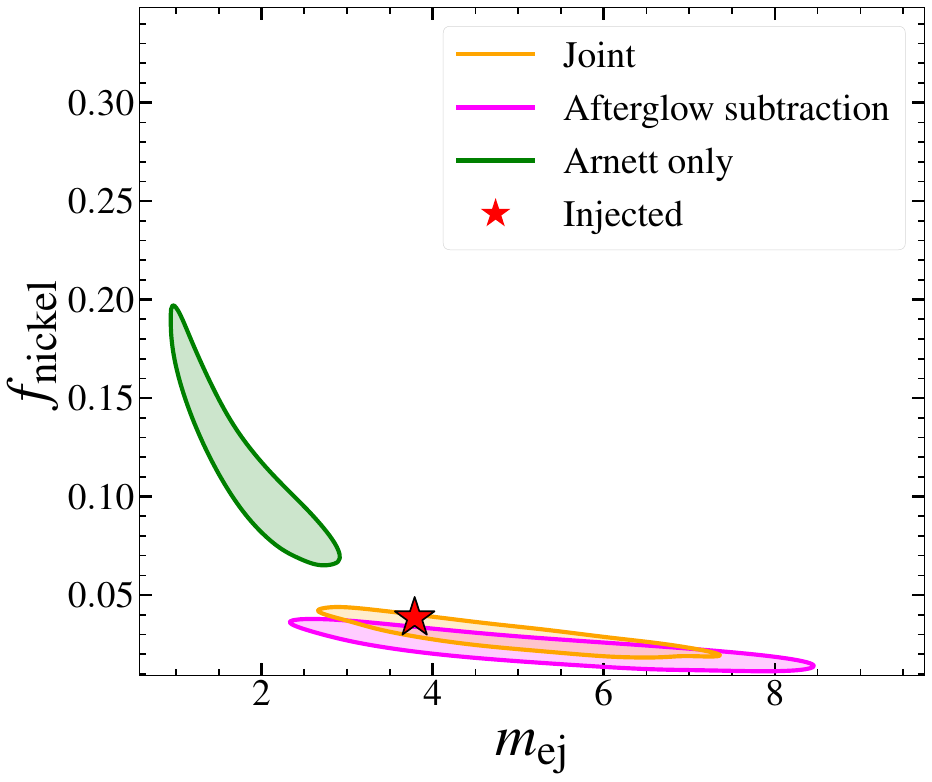}
\caption{Relationship between the $m_\text{ej}$ and $f_\text{nickel}$ parameters for the GRB dominated case with an late jet break using the posteriors obtained from each modelling approach. Each region represents the $1\sigma$ credible interval of the posterior.}
\label{late_sndom_fnimej}
\end{figure}

\begin{figure}[h]
\centering
\includegraphics[width=0.6\linewidth]{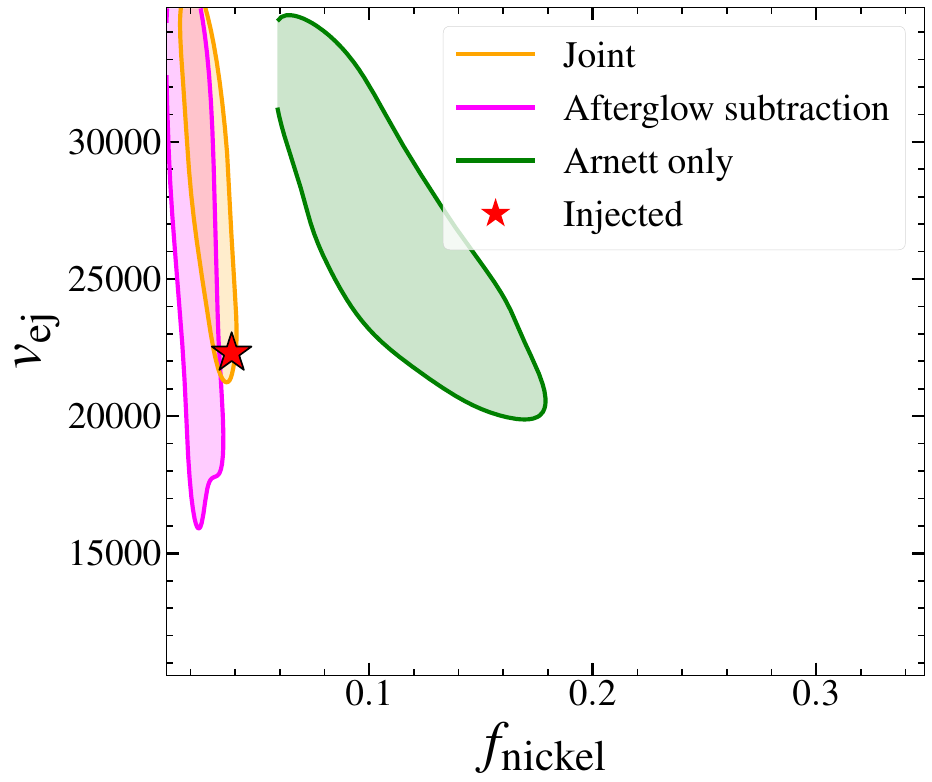}
\caption{Relationship between the $f_\text{nickel}$ and $v_\text{ej}$ parameters for the GRB dominated case with an late jet break using the posteriors obtained from each modelling approach. Each region represents the $1\sigma$ credible interval of the posterior.}
\label{late_sndom_fnimej}
\end{figure}

\begin{figure}[h]
\centering
\includegraphics[width=0.6\linewidth]{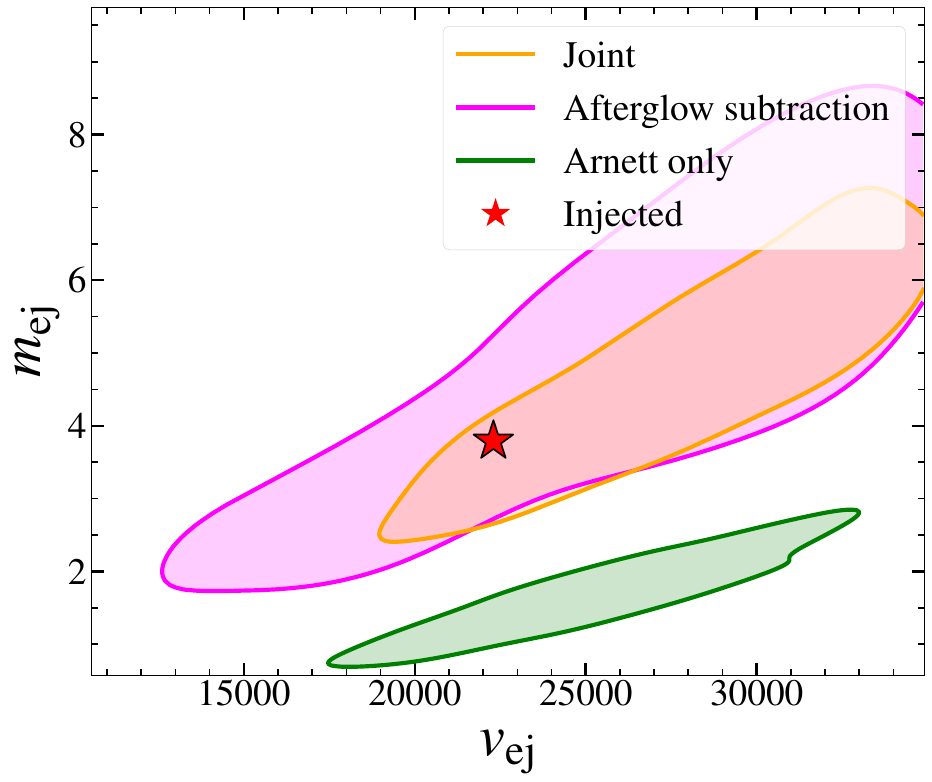}
\caption{Relationship between the $v_\text{ej}$ and $m_\text{ej}$ parameters for the GRB dominated case with an late jet break using the posteriors obtained from each modelling approach. Each region represents the $1\sigma$ credible interval of the posterior.}
\label{late_sndom_vejmej}
\end{figure}

\begin{figure*}[ht!]
\centering
\includegraphics[width=0.58\linewidth]{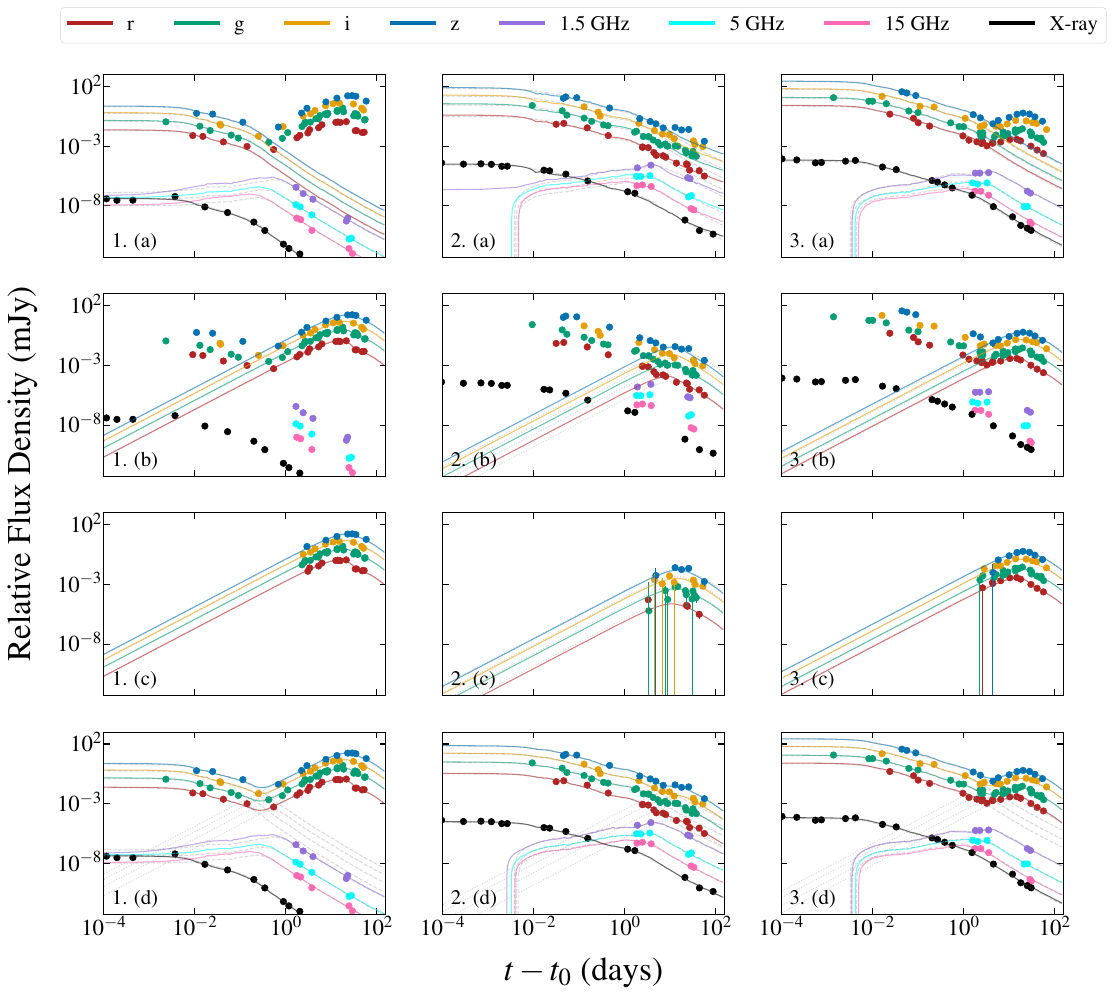}

\caption{Breakdown of fitting results for three chosen cases of GRB-SNe (columns) with late break times, using four different modelling approaches (rows). The columns are the (1) SN-dominated, (2) GRB-dominated, and (3) normal GRB-SN cases, respectively. The rows are the (a)  tophat only model, (b) Arnett only model, (c) afterglow subtracted method and (d) our joint fitting model.The dashed grey lines represent the true GRB light curve and the dotted grey lines represent the true SN light curve produced using the injection parameters. Each band is vertically offset for clarity.}
\label{fitting_all_latetb}

\end{figure*}

\begin{figure*}
\centering
\includegraphics[width=0.58\linewidth]{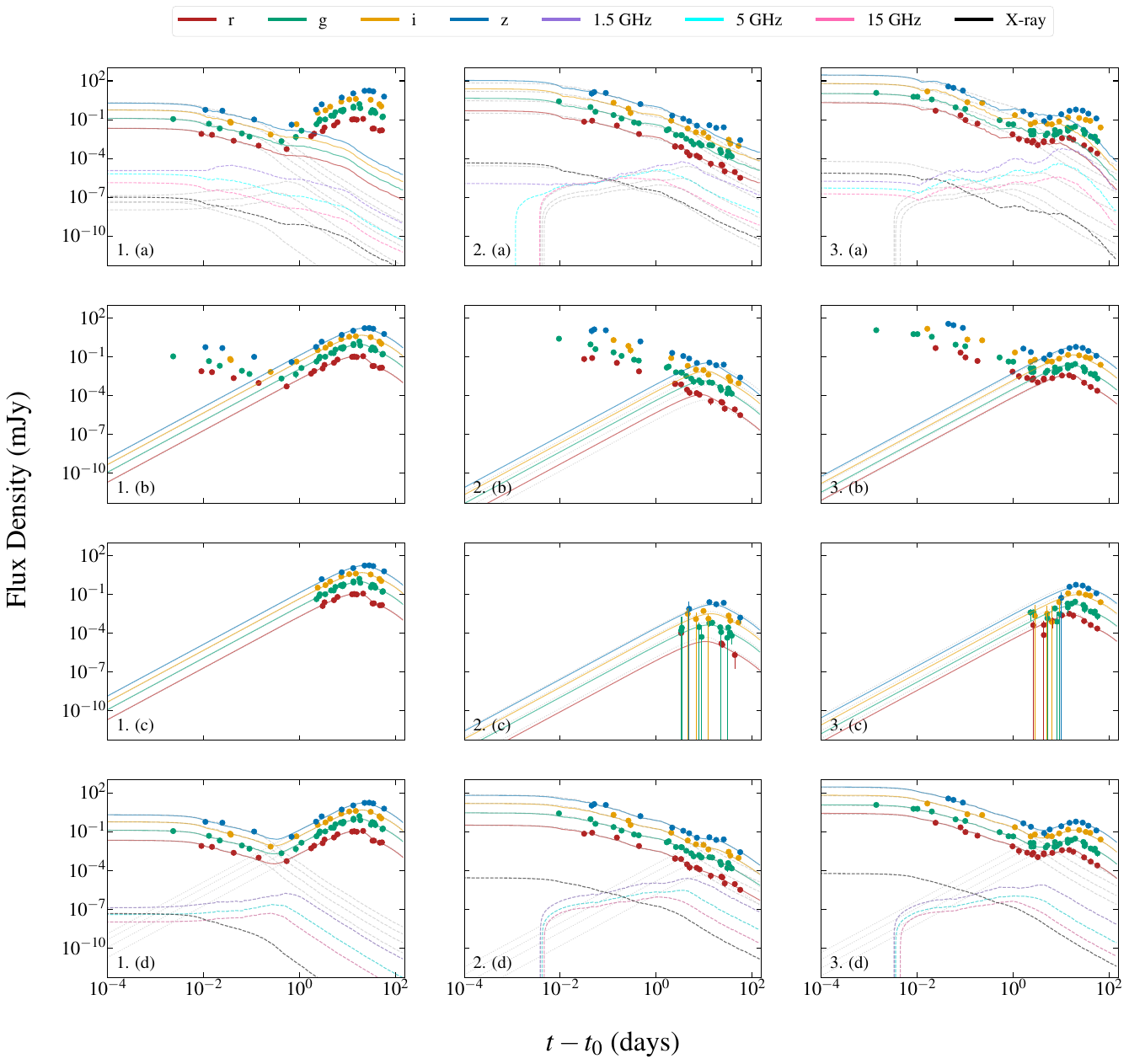}

\caption{Breakdown of fitting results for three chosen cases of GRB-SNe (columns) with late jet break times, using four different modelling approaches (rows) where no X-ray or radio data was used in the modelling. The columns are the (1) SN-dominated, (2) GRB-dominated, and (3) normal GRB-SN cases, respectively. The rows are the (a) top-hat only model, (b) Arnett only model, (c) afterglow subtracted method and (d) our joint fitting model.The dashed grey lines represent the true GRB light curve and the dotted grey lines represent the true SN light curve produced using the injection parameters. Each band is vertically offset for clarity.}
\label{Fitting_no_Xray_latetb}

\end{figure*}

\end{document}